\documentclass[aps,prd,nofootinbib]{revtex4}
\pdfoutput=1

\usepackage{amsmath}
\usepackage{amssymb}
\usepackage{anysize}
\usepackage{bold-extra}
\usepackage{color}
\usepackage{enumerate}
\usepackage{fancyhdr}
\usepackage{graphicx}
\usepackage[linktocpage,colorlinks,urlcolor=blue]{hyperref}
\usepackage{mathrsfs}
\usepackage{multirow}
\usepackage[final]{pdfpages}
\usepackage{rotating}
\usepackage{subfig}
\usepackage{textcomp}
\usepackage{url}
\usepackage{verbatim}
\usepackage{wrapfig}
\usepackage{amsfonts}
\usepackage{comment}
\usepackage{epigraph}
\usepackage[utf8]{inputenc}
\usepackage{slashed}
\usepackage{bbm}
\usepackage{cancel}
\usepackage{epsf, array, color}
\def\nn{\nonumber}

\def\gsim{\mbox{\raisebox{-.6ex}{~$\stackrel{>}{\sim}$~}}}
\begin{document}

\title{Standard Model EFT and  Extended Scalar Sectors}

\author{Sally Dawson}
\email{dawson@bnl.gov}
\affiliation{Department of Physics, Brookhaven National Laboratory, Upton, N.Y., 11973, U.S.A.}

\author{Christopher W. Murphy}
\email{cmurphy@quark.phy.bnl.gov}
\affiliation{Department of Physics, Brookhaven National Laboratory, Upton, N.Y., 11973, U.S.A.}

\begin{abstract}
One of the simplest extensions of the Standard Model is the inclusion of an additional scalar multiplet, and we consider
scalars 
in the $SU(2)_L$ singlet, triplet, and quartet representations.  We 
examine models with heavy neutral scalars, $m_H\sim 1-2$~TeV, and the matching of the UV complete theories to
the low energy effective field theory.  We demonstrate the 
agreement of the kinematic distributions obtained in  the singlet models for the gluon fusion of a Higgs pair
with the predictions of the effective field theory.
The restrictions on the extended scalar sectors due to unitarity and precision electroweak measurements are
summarized and  lead to highly restricted regions of viable parameter space for the triplet and quartet models.
\end{abstract}

\maketitle

\section{Introduction}

The discovery of the Higgs boson at the LHC marks the beginning of the exploration of the nature of 
electroweak symmetry breaking.  Our knowledge of the structure of the scalar potential remains 
primitive -- there are no experimental measurements of the Higgs self-couplings and extended Higgs sectors 
can easily be made  consistent
with LHC data on single Higgs production and searches for heavy neutral
scalars.  The cleanest mechanism for obtaining information on the Higgs tri-linear coupling is a measurement
of double Higgs production from gluon fusion.   In the Standard Model (SM), the rate for double Higgs production is
exceedingly small, presenting a challenge even for the high luminosity LHC.

The simplest extension of the SM scalar sector is the inclusion of an additional scalar multiplet, $\phi$.  If these new scalar 
multiplets, $\phi$,  have light neutral scalars in addition to the $125$~GeV Higgs boson, 
the new scalars  can be studied by direct production and they can also contribute  resonant signatures to double Higgs production.  Alternatively, if 
the new neutral scalars are heavy, $M_\phi \gg m_h$, their contributions to
 low scale physics can be captured in an effective 
field theory framework, with the largest effects coming from dimension-$6$ 
operators~\cite{Kilian:2003xt, Englert:2014uua, Henning:2014wua, deBlas:2014mba, Gorbahn:2015gxa, Brehmer:2015rna, Chiang:2015ura,Carena:2013ooa}.

We concentrate on UV complete models with scalar sectors that  have renormalizable couplings to the SM Higgs
doublet.  The list of such representations is rather short, and the parameters of these models are tightly
restricted by the requirements of perturbative unitarity and agreement with precision electroweak
measurements.  We consider scalars that are $SU(3)_C$ singlets and $SU(2)_L$ singlets, triplets, and
quartets.  (The interesting case of additional $SU(2)_L$ doublets has been extensively 
studied in the literature~\cite{Gunion:2002zf,Khandker:2012zu, Henning:2014wua, deBlas:2014mba, Gorbahn:2015gxa, Chiang:2015ura, Brehmer:2015rna, delAguila:2016zcb, Buchalla:2016bse, Jiang:2016czg,Carena:2013ooa}.)  
Using standard techniques~\cite{Kilian:2003xt,Henning:2014wua}, the heavy $\phi$ can be 
integrated out, leading to predictions for the effective field theory
 dimension-$6$ coefficients corresponding to a given extended scalar model.  We restrict ourselves to
contributions that arise at tree level. Furthermore, we assume no large non-linearities are generated when the heavy $\phi$ is integrated out. That is to say, we are using the SM Effective Field Theory (SMEFT), not Higgs Effective Field Theory (HEFT)~\cite{Feruglio:1992wf, Burgess:1999ha, Grinstein:2007iv}. 
 
 An effective field theory has an expansion in powers of (Energy)$^2/\Lambda^2$, where $\Lambda$ is a
 high scale UV cut-off.  At large values of the energy, kinematic distributions  in the SMEFT can be expected to
 diverge from the exact low energy results~\cite{Contino:2016jqw, Ferreira:2017ymn}.
   A well known example in the case of $gg\rightarrow hh$ is the failure
 of the $m_t\rightarrow\infty$ limit to reproduce the exact cross section and invariant mass
 distribution~\cite{Heinrich:2017kxx,Dolan:2012rv,Chen:2014ask}.  The SMEFT operators
 have different energy dependences in the high energy limit, and kinematic distributions could potentially distinguish
 between the contributions of different dimension-$6$ operators~\cite{Contino:2012xk, Goertz:2014qta, Azatov:2015oxa}.  
 We study the accuracy of the effective
 field theory  for reproducing the predictions of UV complete models  with heavy  scalar singlets
 for the $gg\rightarrow hh$ process and compare the $p_T$ spectrum and the $M_{hh}$ distributions in the SMEFT
 and the UV complete singlet models~\cite{Dolan:2012ac,Bowen:2007ia, Pruna:2013bma, No:2013wsa, Chen:2014ask,deFlorian:2017qfk}.  
 Analogous studies have shown
good agreement for  single Higgs production in selected models with extended scalar sectors~\cite{Gorbahn:2015gxa, Brehmer:2015rna}.

The scalar triplet model violates custodial $SU(2)$ and all of the SMEFT coefficients are proportional to this violation and hence
are forced to be small~\cite{Forshaw:2003kh, SekharChivukula:2007gi, Chen:2008jg, Khandker:2012zu,Englert:2013wga}.  The combined requirements from measurements of the $\rho$ parameter and perturbative 
unitarity lead to models that are indistinguishable from the SM through either single or double Higgs production.  The
quartet models~\cite{AbdusSalam:2013eya} also violate custodial $SU(2)$ and we demonstrate that these models have
extremely restricted  regions of  viable parameter space when perturbative unitarity is enforced.

In Section~\ref{sec:SMEFT}, we review  the framework of the SMEFT with details
given in Appendix~\ref{sec:EFTd}, while Section~\ref{sec:ess} has descriptions of the extended scalar sectors we consider.  Further
details of the models are contained in Appendix~\ref{sec:modD}.  Section~\ref{sec:rho}  contains discussions of limits 
on the parameters of the scalar sectors
from the $\rho$ parameter, single Higgs production, and the requirement of perturbative unitarity.  In particular, a discussion of limits on the SMEFT coefficients in models with extended Higgs sectors from single Higgs production
is given in Section~\ref{sec:singleh}.
The numerical comparison of the SMEFT  and extended scalar models for 
double Higgs production is in Section~\ref{sec:2Higgs}, with conclusions in Section~\ref{sec:con}. 
\section{Standard Model Effective Field Theory}
\label{sec:SMEFT}

The Lagrangian we consider can be written as
\begin{equation}
\label{eq:L}
\mathcal{L}_{SMEFT} = \mathcal{L}_{SM} + \mathcal{L}^{(5)} + \mathcal{L}^{(6)} + \ldots
\end{equation}
where $\mathcal{L}^{(n)}$ has dimension-$n$ and can be parameterized as $\mathcal{L}^{(n)}=\Sigma_i
{c_i^{(n)}\over v^{n-4}}O_i^{(n)}$.

Including only the third generation fermions and neglecting possible mixing with
the lighter generations, the Higgs sector of the SM is given by 
\begin{align}
\label{eq:LSM}
\mathcal{L}_{\text{SM}} \supset \mathcal{L}_H &= \left(D_{\mu} H\right)^{\dagger} \left(D^{\mu} H\right) 
-V_{SM}(H)
\nonumber \\
&- \left[y_b \left(\bar{q}_L b_R H\right) + y_t \left(\bar{q}_L t_R \tilde{H}\right) +y_{\tau} \left(\bar{l}_L \tau_R H\right) + \text{h.c.}\right] \, ,\nonumber
 \\
V_{SM}(H)&=- \mu^2 \left(H^{\dagger} H\right) + \lambda \left(H^{\dagger} H\right)^2 
 \, ,
\end{align}
where $\bar{q}_L$ and $\bar{l}_L$ are the left-handed $(t,b)_L$ and $(\nu,\tau)_L$ doublets, $\tilde {H}\equiv i\sigma_2 H^*$, and the $SU(2)_L$ doublet
$H$ is parameterized as,
\begin{equation}
\label{eq:hdef}
H\equiv 
\begin{pmatrix}H_1 \\
H_2
\end{pmatrix}\equiv
\begin{pmatrix}
w^+ \\
{v_h+h^\prime +iz\over \sqrt{2}}
\end{pmatrix} \, .
\end{equation} 

We are interested in the dimension-$6$  CP-conserving operators generated at tree level by extended scalar sectors,
which takes the form,
\begin{align}
\label{eq:Left}
\mathcal{L}^{(6)} &= \sum_i \frac{c_i}{v^2} O_i \supset \mathcal{L}^{(6)}_H \nn \\
\mathcal{L}^{(6)}_H &= \frac{c_H}{2 v^2} \partial_{\mu}\left(H^{\dagger} H\right)\partial^{\mu} \left(H^{\dagger} H\right) + \frac{c_T}{2 v^2} \left| H^{\dagger} \overleftrightarrow{D}_{\mu} H\right|^2  - \frac{c_6 \lambda}{v^2} \left(H^{\dagger} H\right)^3  \\
&+ \frac{\left(H^{\dagger} H\right)}{v^2} \left[c_b y_b \left(\bar{q}_L b_R H\right) + c_t y_t \left(\bar{q}_L t_R \tilde{H}\right) + c_{\tau} y_{\tau} \left(\bar{l}_L \tau_R H\right) + \text{h.c.}\right] , \nn
\end{align}
where $H^{\dagger} \overleftrightarrow{D}_{\mu} H\equiv H^\dagger D_\mu H-(D_\mu H^\dagger) H$.
In the models we consider  $c_t = c_b = c_{\tau} \equiv c_f$ at tree-level and none of the
extended scalar models we consider generate $H^{\dagger} H V^A_{\mu\nu} V^{A \mu\nu}$ 
 ($V$ is the $SU(3)_C,~SU(2)_L$ or $U(1)$ 
gauge boson) at tree level, so they are not included in Eq.~\eqref{eq:Left}. Minimizing the potential in Eq.~\eqref{eq:LSM} yields the constraint
\begin{equation}
\mu^2 = \lambda v^2\left(1+ \frac{3}{4} c_6\right) ,
\end{equation}
where $v \approx 246$~GeV is the vacuum expectation value (vev) of the Higgs field. With this normalization the coefficients of the operators appearing in Eq.~\eqref{eq:Left} are of order $v^2 / \Lambda^2$, where again $\Lambda$ is the cutoff of the effective theory. In additional to the previously mentioned energy expansion, there is also a mass gap, $v^2 < \Lambda^2$.

We use the basis of Ref.~\cite{Contino:2013kra}, as it has a convenient normalization for our purposes. It is straightforward to convert this basis into a different  one, \textit{e.g.}~\cite{Grzadkowski:2010es}, and Appendix~\ref{sec:EFTd} contains information about operators bases and other SMEFT details. We are primarily interested in the leading order (LO) EFT effects, which generally means dimension-6 operators generated at tree level. 
We note that at one-loop the renormalization group (RG) evolution of the operators $O_H$ and $O_T$ generates operators of the form $\psi^2 H^2 D$~\cite{Jenkins:2013zja, Jenkins:2013wua, Alonso:2013hga}. The subset of dimension-6 operators considered in Eq.~\eqref{eq:Left} is otherwise closed under RG evolution at one-loop.

The kinetic energy for the Higgs boson, $h^\prime$, 
in Eq.~\eqref{eq:L} is not canonically normalized. A field redefinition can be made to correctly normalize
the kinetic energy\footnote{We work to linear order in the coefficients, $c_i$.} and eliminate derivative interactions~\cite{Giudice:2007fh, Buchalla:2013rka} 
\begin{align}
\label{eq:redef}
h^\prime &= h \left[1 -  \frac{c_H}{2} \left(1 + \frac{h}{v} + \frac{h^2}{3 v^2}\right) \right] , \\
\partial_{\mu} h^\prime  &= \partial_{\mu} h \left[1 - \frac{c_H}{2} \left(1 + \frac{h}{v}\right)^2 \right] . \nn
\end{align}
Using Eq.~\eqref{eq:redef} the Higgs boson Lagrangian takes the form
\begin{align}
\label{eq:LeftCan}
\mathcal{L}_h &= \frac{1}{2} \left(\partial_{\mu} h\right)^2 - \frac{1}{2} m_h^2 h^2  - \frac{m_h^2}{2 v} \left(1 + c_6 - \frac{3}{2} c_H\right) h^3 \\
&- \frac{m_h^2}{8 v^2} \left(1 + 6 c_6 - \frac{25}{3} c_H\right) h^4 - \frac{m_h^2}{48 v^4} \left(3 c_6 - 4 c_H\right) h^5 \left(h + 6 v\right) , \nn 
\end{align}
with $m_h \approx 125$~GeV. There are also modifications to the Yukawa sector from Eq.~\eqref{eq:redef}
\begin{equation}
\mathcal{L}_{y_t} = - m_t \bar{t} t \left[1 + \left(1 - \frac{c_H + 2 c_t}{2}\right) \frac{h}{v} - \frac{c_H + 3 c_t}{2} \left(\frac{h^2}{v^2} + \frac{h^3}{3 v^3}\right)\right] ,
\end{equation}
and similarly for the other SM fermions.
\section{Extended Scalar Sectors}
\label{sec:ess}

We consider a number of extensions of the SM where a single new spin-zero multiplet, $\phi$,  is added to the SM and require that 
there is a renormalizable interaction with the SM $H$ doublet that is linear in $\phi$. 
There is a sizable literature on integrating out heavy scalars and studying their SMEFT contributions, see for instance~\cite{Khandker:2012zu, Henning:2014wua, deBlas:2014mba, Gorbahn:2015gxa, Chiang:2015ura, Brehmer:2015rna, delAguila:2016zcb, Buchalla:2016bse, Jiang:2016czg}. The models we consider are: a real singlet ($\mathbf{1}_{0}$), a real triplet ($\mathbf{3}_{0}$), a complex triplet ($\mathbf{3}_{1}$), and two quartets: quartet$_1$ ($\mathbf{4}_{1/2}$) and quartet$_3$ ($\mathbf{4}_{3/2}$). The numbers in parentheses are the  $SU(2)_L \times U(1)_Y$ quantum numbers of the new scalars, all of which are color singlets. These models only generate dimension-6 operators of the form $H^6$ and  $H^4 D^2$ at tree level
(where $D$ is the $SU(2)_L\times U(1)_Y$ covariant derivative). As such they are good candidates to be discovered through deviations in double Higgs production from the SM
predictions. 

The potential can schematically be written as (see also~\cite{AbdusSalam:2013eya})
\begin{equation}
V\left(H, \phi\right) = V_{SM}\left(H\right) + V_{Z_2}\left(H, \phi\right) + V_{\bcancel{Z_2}}\left(H, \phi\right) ,
\end{equation}
where $\phi$ is the new scalar, and $V_{SM}$ is given in  Eq.~\eqref{eq:LSM}.
For a real valued $\phi$, the $Z_2$ preserving potential has the following form
\begin{equation}
V_{Z_2}\left(H, \phi\right) = \frac{1}{2} M^2 \phi^a \phi^a + \lambda_{\alpha} H^{\dagger} H \phi^a \phi^a + \lambda_{\beta} \left(\phi^a \phi^a\right)^2 ,
\end{equation}
where $a$ are the $SU(2)_L$ indices,  and for a complex valued $\phi$ there may be multiple $\alpha$ and/or $\beta$-type
 interactions. Additionally, when $\phi$ is complex, there is no factor of one-half in front of the mass term, and $\phi^a \phi^a$ should be replaced with $\phi^{a \dagger} \phi^a$. Depending on the $SU(2)_L$  representation of $\phi$, the $Z_2$ violating potential contains one of the following interactions
\begin{equation}
 V_{\bcancel{Z_2}} \sim m_1 H^2 \phi \quad \text{or} \quad V_{\bcancel{Z_2}} \sim \lambda_1 H^3 \phi .
 \end{equation}
If $\phi$ is a singlet there can also be a tadpole term and a cubic self-interaction, both of which violate the $Z_2$ symmetry. 
 
The essential features of each model are listed below. Additional details have been relegated to Appendix~\ref{sec:modD}. We define the angle $\alpha$ to characterize the mixing between the neutral, $CP$-even components of $H$ and $\phi$
\begin{equation}
\label{eq:singmix}
\begin{pmatrix}
h \\
{\cal{H}}
\end{pmatrix} = 
\begin{pmatrix}
\cos\alpha & - \sin\alpha \\
\sin\alpha & \cos\alpha
\end{pmatrix}
\begin{pmatrix}
h^\prime \\
\varphi
\end{pmatrix} ,
\end{equation}
where $\text{Re}(H_2) = {v_h + h^\prime\over\sqrt{2}}$ and $\text{Re}(\phi^0) = v_{\phi} + \varphi$, and $v_h$ and $v_{\phi}$ are the vevs of $H$ and $\phi$, respectively. In all of the models we consider, a non-zero value of $\alpha$ leads to a universal modification of the Higgs couplings to SM particles (excluding the Higgs self-couplings). This angle has been bounded
from the single Higgs signal strengths
 by the ATLAS collaboration, with the result $\cos\alpha > 0.94$ at the 95\% confidence level (CL)~\cite{ATLAS:2014kua}.

With the above definitions of the vevs of $H$ and $\phi$, the electroweak (EW) vev is given by
\begin{equation}
v^2 = v_h^2 + 2 \left[t(\phi)\left(t(\phi) + 1\right) - t_3(\phi)^2\right] v_{\phi}^2 ,
\end{equation}
where $t(i)$, $t_3(i)$, and $v_i$ are the representation under $SU(2)_L$ of the $i$th multiplet, the neutral component of the $i$th multiplet, and the vev of the $i$th multiplet, respectively. When $\phi$ is a singlet $v_h = v$, and we define $\tan\beta_s = v_h / v_{\phi}$. For higher $SU(2)_L$ representations, we define the mixing angle between the two vevs as
\begin{equation}
\label{eq:betadef}
v_h = v \cos\beta, \quad v_{\phi} = v \sin\beta / \sqrt{2 \left[t(\phi)\left(t(\phi) + 1\right) - t_3(\phi)^2\right]} .
\end{equation}

The potentials listed below are understood to be in addition to the SM-like potential, $V_{SM}$. Given these interactions, standard methods exist to determine which operators in the SMEFT are generated at tree level in a given model~\cite{Khandker:2012zu, Henning:2014wua}. These results are compiled in Table~\ref{tab:dim6ops}. 
\begin{table}
\centering
 \begin{tabular}{| c | c | c | c | c | c |}
 \hline 
\textbf{Model} & $\mathbf{c_H}$ & $\mathbf{c_6 \lambda_{\textbf{SM}}}$ & $\mathbf{c_T}$ & $\mathbf{c_f}$  \\ \hline 
Real Singlet w/ explicit $\bcancel{Z_2}$ & $\frac{m_1^2 v^2}{M^4}$ & $\frac{m_1^2 v^2}{M^4} \left(\lambda_{\alpha} - \frac{m_1 m_2}{M^2}\right)$ & 0 & 0  \\ \hline
Real Singlet w/ spontaneous $\bcancel{Z_2}$ & $\left(\frac{\lambda_{\alpha} v}{4 \lambda_{\beta} v_{\phi}}\right)^2$ & 0 & 0 & 0  \\ \hline
2HDM & 0 & \checkmark & 0 & \checkmark  \\ \hline
Real Triplet & $- \frac{m_1^2 v^2}{2 M^4}$ & $\frac{m_1^2 v^2 \lambda_{\alpha}}{4 M^4}$ & $ \frac{m_1^2 v^2}{4 M^4}$ & $\frac{m_1^2 v^2}{4 M^4}$    \\ \hline
Complex Triplet & $- \frac{m_1^2 v^2}{2 M^4}$ & $\frac{m_1^2 v^2}{2 M^4} \left(\lambda_{\alpha 1} - \frac{\lambda_{\alpha 2}}{2}\right)$ & $- \frac{m_1^2 v^2}{2 M^4}$ & $\frac{m_1^2 v^2}{2 M^4}$    \\ \hline
Quartet$_1$ & 0 & $- \frac{\lambda_1^2 v^2}{M^2}$ & $\frac{\lambda_1^2 v^4}{2 M^4}$ & 0    \\ \hline
Quartet$_3$ & 0 & $- \frac{\lambda_1^2 v^2}{M^2}$ & $- \frac{3 \lambda_1^2 v^4}{2 M^4}$ & 0    \\ \hline
 \end{tabular}
  \caption{The dimension-6 operators from Eq.~\eqref{eq:Left} that are generated at tree level in the 
  models under consideration. 
  Here $\lambda_{SM} = m_h^2 / 2 v^2$.  
  The 2HDM is listed for the sake of comparison (it additionally generates four-fermion operators).}
  \label{tab:dim6ops}
\end{table}

From Tab.~\ref{tab:dim6ops} we see that taking $\lambda_1$ or $m_1 \to 0$ while holding the other parameters fixed, or sending $M \to \infty$ also while keeping the other parameters fixed, causes the new scalar multiplet to decouple.\footnote{The analogs of $m_1$ and $M$ in the singlet model with spontaneous $Z_2$ breaking are $\lambda_{\alpha}$ and $v_{\phi}$, respectively.} These are the analogs of the alignment without decoupling limit, and  the decoupling limit of the 2HDM, respectively~\cite{Gunion:2002zf, Carena:2013ooa}. 

We also give approximate expressions for the Wilson coefficients in terms of physical masses and mixing angles in Table~\ref{tab:dim6ops_mH}. 
We assume a common mass for the heavy Higgses -- except for $m_A$ ($m_{H^+}$ for the real triplet), which is associated with the alignment without decoupling limit -- and take this mass to be heavy. 
The heavy mass limit needs to be taken with $m_A^2 \sin^2\beta$ fixed for a weakly interacting theory.
Additionally for the triplets and quartets, we assume $\alpha$ is sufficiently small such that it can be neglected.
\begin{table}
\centering
 \begin{tabular}{| c | c | c | c | c | c |}
 \hline 
\textbf{Model} & $\mathbf{c_H}$ & $\mathbf{c_6 \lambda_{\textbf{SM}}}$ & $\mathbf{c_T}$ & $\mathbf{c_f}$  \\ \hline 
Real Singlet w/ explicit $\bcancel{Z_2}$ & $\tan^2\alpha$ & $\tan^2\alpha \left(\lambda_{\alpha} - \frac{m_2}{v} \tan\alpha\right)$ & 0 & 0  \\ \hline
Real Singlet w/ spontaneous $\bcancel{Z_2}$ & $\tan^2\alpha$ & 0 & 0 & 0  \\ \hline
Real Triplet & $- \frac{8 \sin^2\beta\, m_{H^+}^4 }{m_H^4}$ & $\frac{4  \sin^2\beta\, m_{H^+}^6}{m_H^4 v^2}$ & $ \frac{4 \sin^2\beta\, m_{H^+}^4 }{m_H^4}$ & $\frac{4 \sin^2\beta\, m_{H^+}^4 }{m_H^4}$    \\ \hline
Complex Triplet & $- \frac{4 \sin^2\beta\, m_A^4}{m_H^4}$ & $\frac{8 \sin^2\beta\, m_A^6}{m_H^4 v^2}$ & $- \frac{4 \sin^2\beta\, m_A^4}{m_H^4}$ & $\frac{4 \sin^2\beta\, m_A^4}{m_H^4}$    \\ \hline
Quartet$_1$ & 0 & $\frac{24 \tan^2\beta\, m_A^4}{7 m_H^2 v^2}$ & $\frac{24 \tan^2\beta\, m_A^4}{7 m_H^4}$ & 0    \\ \hline
Quartet$_3$ & 0 & $\frac{8 \tan^2\beta\, m_A^4}{3 m_H^2 v^2}$ & $- \frac{8 \tan^2\beta\, m_A^4}{m_H^4}$ & 0    \\ \hline
 \end{tabular}
  \caption{Approximate expressions for the Wilson coefficients in terms of physical masses and mixing angles. We assume a common mass for the heavy Higgses -- except for $m_A$ ($m_{H^+}$ for the real triplet), which is associated with the alignment without decoupling limit -- and take this mass to be heavy. Additionally for the triplets and quartets, we assume $\alpha$ is sufficiently small such that it can be neglected.}
  \label{tab:dim6ops_mH}
\end{table}

\begin{itemize}
\item \textbf{Singlets:} The most general renormalizable potential is
\begin{equation}
\label{eq:Vsing}
V_s = \tfrac{1}{2} M^2 \phi^2 + m_1 H^{\dagger} H \phi + m_2 \phi^3 + m_3^3 \phi + \lambda_{\alpha} H^{\dagger} H \phi^2 + \lambda_{\beta} \phi^4 .
\end{equation}
If $m_{1,2,3}\rightarrow 0$, the potential exhibits an explicit $Z_2$ symmetry.  In the absence of a $Z_2$ 
symmetry, the parameters can be redefined to eliminate a vev for $\phi$.
In terms of the masses of the Higgs bosons and the mixing angle $\alpha$, the Wilson coefficient $c_H$ is the same 
whether or not there is an explicit $Z_2$ symmetry,
\begin{equation}
\label{eq:cHmassES}
c_H = \frac{\left(m_H^2 - m_h^2\right)^2 \sin^2 2\alpha}{\left(m_H^2 + m_h^2 + \left(m_H^2 - m_h^2\right) \cos2\alpha\right)^2} .
\end{equation}
The limiting forms of Eq.~\eqref{eq:cHmassES} are
\begin{equation}
\label{eq:cHmasslimit}
c_H = 
  \begin{cases}
    \tan^2\alpha       & \quad m_H \to \infty \\
    \left(1 - \frac{m_h^2}{m_H^2}\right)^2 \alpha^2  & \quad \alpha \to 0 \\
  \end{cases}
\end{equation}
When the EFT coefficients are expressed in terms of the mass eigenstate parameters, we see from Eq.~\eqref{eq:cHmasslimit} that sending $\alpha \to 0$ is equivalent to taking the alignment without decoupling limit, but sending $m_H \to \infty$ is not the same as taking the decoupling limit.
   \begin{itemize}
   \item \textbf{Real Singlet with Explicit $Z_2$ Breaking:}
When the $Z_2$ symmetry for $\phi$ is explicitly broken, the parameters in Eq.~\eqref{eq:Vsing} can be redefined such that $\phi$ does not get a vev. Parameter space exists such that this the deepest of the possible vacua in
 the theory~\cite{Chen:2014ask,Espinosa:2011ax}. In addition, we redefine $M$ to allow for an easier comparison with the spontaneous symmetry breaking case. The potential is then,
\begin{equation}
\label{eq:Vsing}
V_s = \tfrac{1}{2} M^2 \phi^2 + m_1 \left(H^{\dagger} H - \tfrac{v^2}{2}\right) \phi + m_2 \phi^3 + \lambda_{\alpha} \left(H^{\dagger} H - \tfrac{v^2}{2}\right) \phi^2 + \lambda_{\beta} \phi^4 .
\end{equation} 
In this case, 
\begin{equation}
\lambda_{SM}c_6=c_H\biggl(\lambda_\alpha-{m_1m_2\over M^2}\biggr)\, ,
\end{equation} 
and $\lambda_\alpha$
and $m_2$ are free parameters limited by perturbative unitarity, precision electroweak measurements, and the 
minimization of the potential, while $m_1$ can be expressed in terms of $m_h,~m_H$ and $\alpha$.  
For large $M$, $M\sim m_H$. 
   \item \textbf{Real Singlet with Spontaneous $Z_2$ Breaking:}
In the case of an explicit $Z_2$ symmetry, $\phi$ develops a vev, $\phi = v_{\phi} + \varphi$.  This spontaneously breaks the symmetry, and leads to the following potential:
\begin{equation}
\label{eq:VsingZ2}
V_s = \lambda_{\alpha} \left(H^{\dagger} H - \tfrac{v^2}{2}\right)  \left(\phi^2 - v_{\phi}^2\right) + \lambda_{\beta} \left(\phi^2 - v_{\phi}^2\right)^2 .
\end{equation}
In this scenario $c_6$ vanishes at tree level due to the explicit $Z_2$ symmetry, but $c_H$ is still non-zero~\cite{Gorbahn:2015gxa}.
   \end{itemize}
\item \textbf{Triplets:} We use an adjoint notation for the triplets \begin{equation}
\label{eq:tripadj}
\phi = \phi^a T^a = \frac{1}{2} 
\begin{pmatrix}
\phi^Y & \sqrt{2} \phi^{Y+1} \\
\sqrt{2} \phi^{Y-1} & - \phi^Y
\end{pmatrix} ,
\end{equation}
where $Y$ is the hypercharge of the triplet, and $T^a = \sigma^a / 2$ with $\sigma^a$ being the Pauli matrices. All of the Wilson coefficients in the triplet models are proportional to $c_T$, indicating there is limited potential for these models to modify double-Higgs production since $c_T$ is constrained by the $\rho$ parameter, see Eq.~\eqref{eq:rhofit}. 
   \begin{itemize}
   \item \textbf{Real Triplet:} The real $SU(2)_L$ triplet is hypercharge neutral. The potential in this case is
\begin{equation}
\label{eq:Vrealtrip}
V_{t_r} =  \tfrac{1}{2} M^2 \phi^a \phi^a + m_1 H^{\dagger} T^a H \phi^a + \lambda_{\alpha} H^{\dagger} H \phi^a \phi^a + \lambda_{\beta} \left(\phi^a \phi^a\right)^2 .
\end{equation}
Using  Eq. \eqref{eq:rhodef}, the  Wilson coefficients can be seen to all be proportional to
$\rho-1$:  
\begin{equation}
c_T=c_f=-{c_H\over 2}=\rho-1,\quad  c_6\lambda_{SM}=(\rho-1)\lambda_\alpha\, .
\end{equation}
   \item \textbf{Complex Triplet:} The complex triplet has hypercharge one. Much of the discussion is similar to the real case. The potential is
\begin{align}
\label{eq:Vcomptrip}
V_{t_c} &=  M^2 \phi^{\dagger a} \phi^a + m_1 \left(H^{\dagger} T^a \tilde{H} \phi^a + \text{ h.c.} \right) + \lambda_{\alpha1} H^{\dagger} H \phi^{\dagger a} \phi^a \\
&+ i \lambda_{\alpha2} H^{\dagger} T^a H \epsilon^{abc} \phi^{\dagger b} \phi^c + \lambda_{\beta1} \left(\phi^{\dagger a} \phi^a\right)^2 - \lambda_{\beta2} \epsilon^{abc} \epsilon^{ade}  \phi^{\dagger b} \phi^c \phi^{\dagger d} \phi^e . \nn
\end{align}
   \end{itemize}
   The relations between the coefficients in the complex triplet case are different than in the real case, but again
   all of the Wilson coefficients are proportional to $\rho-1$:
   \begin{equation}
   c_T=-c_f=c_H=\rho-1,\quad  c_6\lambda_{SM}=(1-\rho)\biggl(\lambda_{\alpha_1}-
   {\lambda_{\alpha_2}\over 2}\biggr)\, .
   \end{equation}

\item \textbf{Quartets:} The $SU(2)_L$ quartets of interest have either hypercharge $Y = 3 / 2$ or $1 / 2$. In both cases, the $Z_2$ preserving part of the potential is
\begin{align}
\label{eq:VquarZ2}
V_{q,Z_2} &= M^2 \phi^{* ijk} \phi_{ijk} + \lambda_{\alpha1} H^{* i} H_i \phi^{*ljk} \phi_{ljk} + \lambda_{\alpha2} H^{* i} H_k \phi^{*ljk} \phi_{lji} \\
&+ \lambda_{\beta1} \left(\phi^{* ijk} \phi_{ijk}\right)^2 + \lambda_{\beta2} \phi^{* ijk} \phi_{ijn} \phi^{* lmn} \phi_{lmk} . \nn
\end{align}
We use a symmetric tensor notation, $\phi = \phi_{(ijk)}$~\cite{Hisano:2013sn}, where the
indices are summed over. Since the Young's Tableau for $SU(2)$ only has one row, and representations are symmetric with respect to exchange of blocks of a given row, a $2j+1$ $SU(2)$ multiplet can be written as a $2j$ index symmetric tensor.
   \begin{itemize}
   \item \textbf{Quartet$_1$:} In the $Y = 1 / 2$ case, the $Z_2$ breaking term is
\begin{equation}
\label{eq:Vquar1}
V_{q_1,{\bcancel{Z_2}}} = - \lambda_1 \left(\phi^{* ijk} H_i H_j \epsilon_{k l} H^{* l} + \text{ h.c.}\right) .
\end{equation}
The only dimension-6 operator generated is~\cite{Henning:2014wua}
\begin{equation}
c_6 \lambda_{SM}= - \frac{\lambda_1^2 v^2}{M^2} .
\end{equation}
The quartet is the only model considered  here that contains cubic interactions of the SM $H$ doublet with $\phi$, 
leading to dimension-$6$ coefficients of ${\cal{O}}({1\over M^2})$.
The same value for $c_6$ is generated in the $Y=3/2$ case.
Once EW symmetry is broken $c_T$ is generated at tree level through a dimension-8 operator, see Sec.~\ref{sec:rho}.
When $H$ gets a vev, $\phi$ is forced to get a vev. 
Using the results of Appendix~\ref{sec:modD} we find
\begin{equation}
v_{\phi} \approx \frac{\sqrt{3} \lambda_1 v_h^3}{6 M^2} .
\end{equation}
The vev of $\phi$  leads to the  contribution to $c_T$,
\begin{equation}
c_T=-{1\over 2}{v^2\over M^2} (c_6\lambda_{SM})\, .
\end{equation}
   \item \textbf{Quartet$_3$:} In the $Y = 3 / 2$ case, the $Z_2$ breaking part of the potential is
\begin{equation}
\label{eq:Vquar3}
V_{q_3{\bcancel{Z_2}}} = - \lambda_1 \left(\phi^{* ijk} H_i H_j H_k + \text{ h.c.}\right) .
\end{equation}
In this case the vev of $\phi$ is
\begin{equation}
v_{\phi} \approx \frac{\lambda_1 v_h^3}{2 M^2} \, ,
\end{equation}
leading to 
\begin{equation}
c_T={3\over 2}{v^2\over M^2} (c_6\lambda_{SM})\, .
\end{equation}
   \end{itemize}
\end{itemize}
 
The set of  models considered in this work only generate dimension-6 operators of the form $H^6$, $H^4 D^2$ at tree level. This is not obvious from Tab.~\ref{tab:dim6ops} because we are using a non-redundant set of operators. An additional scalar operator, $O_R$, is generated by some of the models. However, when $O_R$ is eliminated from our operator basis, see Appendix~\ref{sec:EFTd}, operators of the form $\psi^2 H^3$ are generated in addition to purely scalar operators. In contrast with the models we consider, the Two-Higgs Doublet Model (2HDM) generically leads to operators of the form $\psi^2 H^3$, $\psi^4$, even if redundant operators are retained.\footnote{The complex triplet can also interact with SM fermions. In particular, there could be the lepton number violating interaction, $\bar{\ell}_L \phi^{\dagger} (i \sigma^2 \ell^c_L) + \text{h.c.}$ We assume the Yukawa coupling associated with this interaction is negligibly small, consistent with the existence of tiny neutrino masses.}  Due to this complication, and the fact that the 2HDM is extremely well studied, we do not analyze it in this work. See Refs.~\cite{Henning:2014wua, deBlas:2014mba, Gorbahn:2015gxa, Brehmer:2015rna, Belusca-Maito:2016dqe, Jiang:2016czg} for some studies of the 2HDM in an effective field theory context. 

For the singlet model, we separately considered the cases of  explicit and spontaneous $Z_2$ symmetry breaking. What happens when a $Z_2$ symmetry is imposed on a triplet or higher representation? If $\phi$ gets a vev, there is a leftover global $U(1)$ symmetry that leaves the $CP$-odd Higgs boson massless.\footnote{In the real triplet model, it is the charged Higgs boson that is massless in this scenario.} There are two ways out of this problem. The first solution is to not allow the additional multiplet to get a vev. This is the analog of the inert 2HDM~\cite{Deshpande:1977rw}. In this case, no dimension-6 operators are generated at tree level, both in the inert 2HDM and in the higher representation models as well. Alternatively, the pseudoscalar will acquire a mass in the higher representation models if the $Z_2$ symmetry is softly broken. In the triplet models it is possible to achieve a soft breaking of the $Z_2$ symmetry, just as in the 2HDM, through the interaction with coefficient $m_1$. This is not the case for the quartet models where the only (renormalizable) $Z_2$ violating interaction is marginal.
\section{Constraints}
\label{sec:rho}
\subsection{The Rho Parameter}
The $\rho$ parameter is defined as the ratio of neutral to charged currents at low energies~\cite{Ross:1975fq}
\begin{equation}
\rho = 
{M_W^2\over M_Z^2 \cos^2\theta_W} .
\end{equation}
A recent global fit to EW precision data yielded the value~\cite{Erler:2017vaq}
\begin{equation}
\label{eq:rhofit}
{\rho}_{\text{exp.}} = 1 + \left(3.6 \pm 1.9\right) \cdot 10^{-4} . 
\end{equation}
In terms of dimension-6 operators, the $\rho$ parameter takes the form
\begin{equation}
\label{eq:rho6}
\rho = 1 + c_T .
\end{equation}
Alternatively, the tree level contribution in the extended scalar models can be written in terms of the Higgs vevs~\cite{Olive:2016xmw}
\begin{equation}
\label{eq:rhoPDG}
\rho = \frac{\sum_i \left[t(i) \left(t(i )+1\right) - t_3(i)^2\right] v_i^2}{2 \sum_i t_3(i)^2 v_i^2} .
\end{equation}
From Eq.~\eqref{eq:rhoPDG} we see that the $\rho$ parameter generally differs from one in theories with triplets or quartets. The numerator of Eq.~\eqref{eq:rhoPDG} is equivalent to $v^2 / 2$ (with $v \approx 246$~GeV). We  use this fact to eliminate one term from the sum in Eq.~~\eqref{eq:rhoPDG}, say the $i  = 1$ term. If the $i = 1$ $SU(2)_L$ multiplet is taken to be a doublet, possibly SM-like, Eq.~\eqref{eq:rhoPDG} simplifies to
\begin{equation}
\label{eq:rhoPDGs}
\rho = \frac{v^2}{v^2 - 2 \sum_{i > 1}\left[t(i)  \left(t(i )+1\right) - 3 t_3(i)^2\right] v_i^2 } .
\end{equation}

We can compare the calculations of $\rho$ in the unbroken and broken phases of the theories, Eqs.~\eqref{eq:rho6} and~\eqref{eq:rhoPDGs}, respectively. Using the results of Appendix~\ref{sec:modD} we have checked that  for the triplet models, with the reasonable approximations $v_{\phi} \ll v$ and $v_h \approx v$, the calculations of $\rho$ agree in the two different phases. 

In terms of the mixing angle $\beta$, the tree level contribution of each model to the $\rho$ parameter is given in Table~\ref{tab:rho}. Also shown in Tab.~\ref{tab:rho} is the bound on $\beta$ from Eq.~\eqref{eq:rhofit}. Since the global fit prefers a value for $\rho$ slightly greater than one, the models that contribute positively to $\rho$ are somewhat less constrained than those that contribute negatively to $\rho$.
\begin{table}
\centering
 \begin{tabular}{| c | c | c |}
 \hline 
\textbf{Model} & $\mathbf{\rho}$ & \textbf{$3\sigma$ upper limit on $\beta$}  \\ \hline 
Singlet & 1 & none \\ \hline
2HDM & 1 & none \\ \hline
Real Triplet & $\sec^2\beta$ & 0.030  \\ \hline
Complex Triplet & $2\left(3 - \cos2\beta\right)^{-1}$ & 0.014   \\ \hline
Quartet$_1$ & $7\left(4 + 3 \cos2\beta\right)^{-1}$ & 0.033  \\ \hline
Quartet$_3$ & $\left(2 - \cos2\beta\right)^{-1}$ & 0.010  \\ \hline
 \end{tabular}
  \caption{The tree level contribution to $\rho$ in a given model, and the corresponding $3\sigma$ upper limit on the mixing angle $\beta$.}
  \label{tab:rho}
\end{table}

The preceding discussion makes it clear that from an experimental point of view the vevs of the triplets and the quartets are equally well constrained. On the other hand, as mentioned in Sec.~\ref{sec:ess}, from an effective field theory point of view the origins of these vevs are different.  
Since experimentally $\rho$ is measured at low energies, one must consider the effects of the EW symmetry breaking. Given the following dimension-8 operators,
\begin{equation}
\mathcal{L}^{(8)} \supset \frac{c_{H8}}{v^4} \left(H^{\dagger} H\right) \partial_{\mu}\left(H^{\dagger} H\right)\partial^{\mu} \left(H^{\dagger} H\right) + \frac{c_{T8}}{v^4} \left(H^{\dagger} H\right) \left| H^{\dagger} \overleftrightarrow{D}_{\mu} H\right|^2 ,
\end{equation}
once $H$ gets a vev there is an additional contribution to the $\rho$ parameter\footnote{It is well known that the $\rho$ parameter is equivalent to the $T$ parameter of Peskin and Takeuchi~\cite{Peskin:1991sw}. Note however that the operator with Wilson coefficient $c_{T8}$ is not equivalent to the $U$ parameter, which is an $H^4D^4$ operator not an $H^6D^2$ operator.}
\begin{equation}
\label{eq:rhodef}
\rho = 1 + c_T + c_{T8} .
\end{equation}
It is really $c_{T8}$ that is generated in the quartet models, not $c_T$ as shown in Table~\ref{tab:dim6ops}, but from a low energy perspective there is no practical difference.

As previously mentioned, all of the dimension-6 operators generated at tree level in the triplet models are proportional to $c_T$, which constrains the size of those Wilson coefficients to be small. This is not the case in the quartet models, which is a result of the fact that $c_6, c_T$, and ~$c_{T8}$ are generated at different orders in the EFT expansion.
\subsection{Single Higgs Production}
\label{sec:singleh}
Quite generically, theories that modify the rate for double Higgs boson production will also modify the production rate for a single 125~GeV Higgs boson, as well as the Higgs boson's branching ratios. For the models  with extended
scalar sectors that we are interested in, measurements of the 125 GeV Higgs boson yield the bound $\cos\alpha > 0.94$ at the 95\%~CL~\cite{ATLAS:2014kua}. This suppresses the production of the heavy neutral Higgs boson by $\sin^2\alpha$ with respect to the SM rate, which is below the current experimental sensitivities~\cite{Chalons:2016jeu}. We are interested in bounding the Wilson coefficients  that affect single Higgs production, $c_H$ and $c_f$ that are generated in the extended scalar models. The fit is particularly simple in these models, since other
potential dimension-$6$ operators affecting Higgs couplings  are not generated.

For a given Higgs boson production and decay process, $i \to h \to f$, the signal strength is defined as
\begin{equation}
\mu_i^f = \mu_i \cdot \mu^f = \frac{\sigma(i \to h)}{\left(\sigma(i \to h)\right)_{\text{SM}}} \cdot \frac{\text{Br}(h \to f)}{\left(\text{Br}(h \to f)\right)_{\text{SM}}} .
\end{equation}
We use the combined results of ATLAS and CMS based on 7 and 8~TeV data~\cite{Khachatryan:2016vau}, which can be found in Table~\ref{tab:signalstrengths}. The three leftmost columns of Tab.~\ref{tab:signalstrengths} are adapted from Ref.~\cite{Cacchio:2016qyh}, which obtains the values of the signal strengths from Table~13 of~\cite{Khachatryan:2016vau}, and estimates the correlations between the signal strengths from Figure~14 of~\cite{Khachatryan:2016vau}. The rightmost column is  the signal strength in the SMEFT for the operators in Eq.~\eqref{eq:Left}.
For loop level processes, we use the approximate expressions for the signal strengths given in Ref.~\cite{Grinstein:2013npa}. SM Higgs boson branching ratios
and the total width are taken from Ref.~\cite{Heinemeyer:2013tqa}.
\begin{table}
\begin{tabular}{| c | c | c c | c |}
\hline
\textbf{Signal strength} & \textbf{Value} & \multicolumn{2}{|l|}{\textbf{Correlation matrix}} & $\mathbf{\mu}_{\textbf{SMEFT}}$\\
\hline
$\mu_\text{ggF+tth}^{\gamma \gamma}$ & $1.16\pm 0.26$  & 1 & $-0.30$ & $1 - c_H + 0.01 c_f$ \\[3pt] 
$\mu_\text{VBF+Vh}^{\gamma \gamma}$ & $1.05\pm 0.43$  & $-0.30$ & 1 & $1 - c_H + 2.01 c_f$ \\[3pt] 
\hline
$\mu_\text{ggF+tth}^{bb}$ & $1.15\pm 0.97$  & $1$ & $4.5\cdot 10^{-3}$ & $1 - c_H + 2.55 c_f$ \\[3pt] 
$\mu_\text{VBF+Vh}^{bb}$ & $0.65\pm 0.30$  & \:$4.5\cdot 10^{-3}$ & $1$ & $1 - c_H + 0.55 c_f$ \\[3pt] 
\hline
$\mu_\text{ggF+tth}^{\tau \tau}$ & $1.06\pm 0.58$  & 1 & $-0.43$ & $1 - c_H + 2.55 c_f$ \\[3pt] 
$\mu_\text{VBF+Vh}^{\tau \tau}$ & $1.12\pm 0.36$  & $-0.43$ & 1 & $1 - c_H + 0.55 c_f$ \\[3pt] 
\hline
$\mu_\text{ggF+tth}^{WW}$ & $0.98\pm 0.21$  & 1 & $-0.14$ & $1 - c_H + 0.55 c_f$ \\[3pt]  
$\mu_\text{VBF+Vh}^{WW}$ & $1.38\pm 0.39$  & $-0.14$ & 1 & $1 - c_H + 1.45 c_f$ \\[3pt] 
\hline
$\mu_\text{ggF+tth}^{ZZ}$ & $1.42\pm 0.35$  & 1 & $-0.49$ & $1 - c_H + 0.55 c_f$ \\[3pt]  
$\mu_\text{VBF+Vh}^{ZZ}$ & $0.47\pm 1.37$  & $-0.49$ & 1 & $1 - c_H + 1.45 c_f$ \\[3pt] 
\hline
\end{tabular}
\caption{Higgs boson signal strengths from~\cite{Khachatryan:2016vau}. The right column 
has  the signal strengths in the SMEFT for the operators in Eq.~\eqref{eq:Left}. The left three columns are adapted from~\cite{Cacchio:2016qyh}.}
\label{tab:signalstrengths}
\end{table} 

The method of least squares is used to find the favored parameter space. The $\chi^2$ function schematically is
\begin{equation}
\chi^2 \sim \left(\vec{\mu}_i^f - \vec{\mu}_{\text{SMEFT}}\right)^{\top} V^{-1} \left(\vec{\mu}_i^f - \vec{\mu}_{\text{SMEFT}}\right) ,
\end{equation}
where $V$ is the covariance matrix of the experimental values. The parameter values that minimize $\chi^2$ are 
\begin{align}
\label{eq:eftlim}
10^2 c_H &= - 8.8 \pm 9.9 , \quad 10^2 c_f = 5.0 \pm 10.7  \\
r &= 
\begin{pmatrix}
1 & - 0.196 \\
-0.196 & 1
\end{pmatrix} , \nn
\end{align}
where the correlation matrix is denoted $r$ to avoid confusion with the $\rho$ parameter. 
Parameterizing the SMEFT coefficients as $c_H\sim {\hat c}_H {v^2\over M^2}$,  
 $c_f\sim {\hat{c}}_f {v^2\over M^2}$, the $95\%$ confidence level limits from 
Eq.~\eqref{eq:eftlim} are,
\begin{eqnarray}
\label{eq:clims}
-18\biggl({M\over 2~\text{TeV}}\biggr)^2< &{\hat c}_H& < 7 \biggl({M\over 2~\text{TeV}}\biggr)^2\nonumber \\
-2\biggl({M\over 2~\text{TeV}}\biggr)^2< &{\hat c}_f& < 5 \biggl({M\over 2~\text{TeV}}\biggr)^2
\, . \nonumber \\
\end{eqnarray}
The SMEFT coefficients predicted in the
previous sections from the extended scalar sectors are comfortably within the limits of Eq.~\eqref{eq:clims}.

\begin{figure}
  \centering
\includegraphics[width=0.5\textwidth]{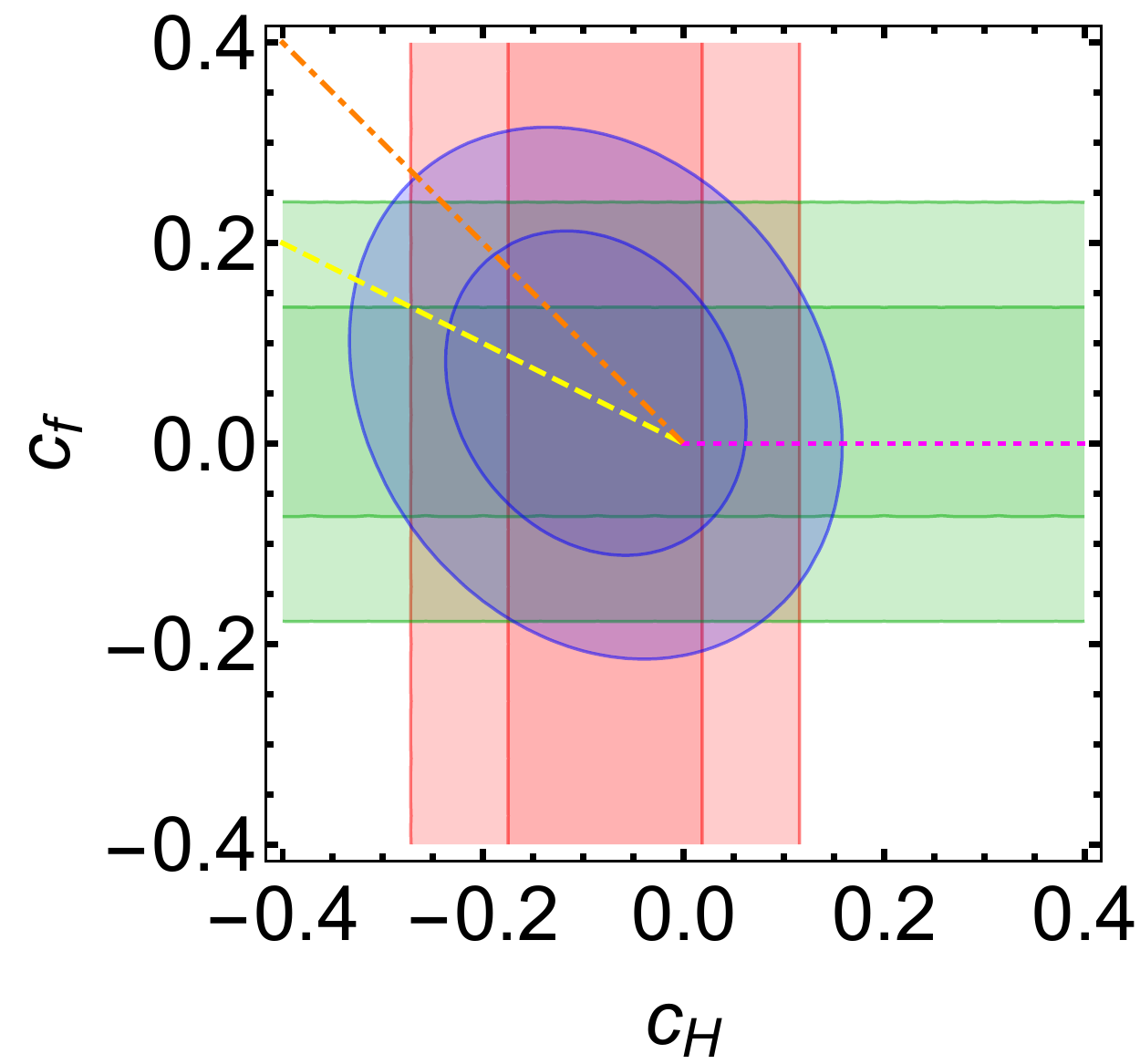}
 \caption{Results of  a $\chi^2$ fit to Higgs data to limit $c_H$ and $c_f$. The darker and lighter regions represent the $1\sigma$ and $2\sigma$ confidence regions, respectively. The magenta (dotted), yellow (dashed),  and 
 orange (dot-dashed) line segments correspond to the real singlet, real triplet, and complex triplet models, respectively. The signs of $c_H$ and $c_f$ are fixed in these models, which is why the line segments do not cover the whole plane.}
   \label{fig:cHcf}
\end{figure}

The confidence regions for the estimated parameters are determined using $\chi^2 \leq \chi^2_{\text{min}} + \Delta \chi^2$, where the $1\sigma$ and $2\sigma$ regions are given by $\Delta \chi^2 = 1, 4\, (2.30, 6.18)$ when the number of parameters to be estimated is 1 (2). The results of this fit are shown in  Fig.~\ref{fig:cHcf}. The darker and lighter regions represent the $1\sigma$ and $2\sigma$ confidence regions, respectively. The red and green regions are fits to $c_H$ or $c_f$ with the other parameter fixed to zero, while the blue region is a simultaneous fit to both parameters. Also shown in Fig.~\ref{fig:cHcf} are the predictions for the real singlet (magenta, dotted), real triplet (yellow, dashed), and complex triplet (orange,  dot-dashed) models imposing the
relations between coefficients shown in Tab.~\ref{tab:dim6ops}. The signs of $c_H$ and $c_f$ are fixed in these models, which is why the line segments do not cover the whole plane.
\subsection{Perturbative Unitarity}

There are a number of theoretical considerations that can be used to constrain the parameter space of the extended scalar sectors, including requiring the potential to be bounded from below, or requiring the EW vev to be the deepest of the vacua in the theory.\footnote{General bounded from below conditions for models of the type we are interested in can be found in Ref.~\cite{Kannike:2016fmd}.} In this work we focus on theoretical constraints coming from perturbative unitarity~\cite{Lee:1977eg}. In non-renormalizable theories, such as the SMEFT, scattering amplitudes generally grow with energy leading to a breakdown of unitarity at some critical energy. On the other hand, the extended scalar sectors under consideration are unitarity, and their $2 \to 2$ scattering amplitudes do not grow with energy at large $s$. However, the same approach may still be used to examine where the breakdown of perturbation theory occurs. If a certain combination of parameters appearing in a scattering amplitude is too large, the tree level amplitude will not be a good approximation of the full amplitude.

Our approach to finding the unitarity or perturbativity bounds is the same in both cases. We compute all the $2 \to 2$ scattering amplitudes, $\mathcal{M}_{i \to f}(s,t)$, in the scalar sector of a given theory, including those containing Goldstone bosons. The set of initial states in the SM  and SMEFT with a net electric charge of zero, for example, is $i = \{w^+ w^-,\, z z,\, h h,\, h z\}$. The computation is done in the limit that the center-of-mass energy, $\sqrt{s}$, is much larger than the other scales in the problem. For renormalizable theories this simplifies the scattering amplitudes to a linear of combination of quartic couplings. The matrix of $\ell = 0$ partial-wave amplitudes, $\left(a_0\right)_{i, f}$, is then computed from these scattering amplitudes
\begin{equation}
\left(a_0\right)_{i, f} = \frac{1}{16 \pi s} \int_{-s}^0 \! dt \, \mathcal{M}_{i \to f}(s,t) .
\end{equation}
The eigenvalues, $a_0$, of this matrix are bounded by the unitarity of the $S$-matrix
\begin{equation}
\label{eq:a0re}
\left|\text{Re}\left(a_0\right)\right| \leq \frac{1}{2} .
\end{equation}
The unitarity or perturbativity bounds derived in this work ultimately come from~\eqref{eq:a0re}. For a point in the parameter
space to be considered viable, we require that Eq.~\eqref{eq:a0re} is satisfied for every eigenvalue for that choice of parameters unless otherwise specified. 

We begin by discussing the unitarity bounds on the SMEFT. The Feynman rules for the SMEFT in $R_{\xi}$ gauge have recently been presented in Ref.~\cite{Dedes:2017zog}. Using the results of~\cite{Dedes:2017zog}, and neglecting terms that do not grow with energy, we find the (unique) eigenvalues of the matrix of partial-wave amplitudes for high energy scalar scattering in the SMEFT are
\begin{equation}
a_0 = \frac{s}{32 \pi v^2} \{3 \left(c_T + c_H\right),\, 3 c_T - c_H,\, - \left(3 c_T + c_H\right),\, - c_H\}
\end{equation}
Since $c_T$ is constrained to be small by the $\rho$ parameter, we can ignore it in determining the critical energy. Our results are in agreement with Ref.~\cite{Giudice:2007fh}, which considered a subset of amplitudes (and only $c_H$). With these approximations, we find the SMEFT will break down at an energy no higher than
\begin{equation}
E_{\text{crit.}} \approx \sqrt{\frac{16 \pi}{3 c_H}} v \approx \frac{1~\text{TeV}}{\sqrt{c_H}} .
\end{equation}

\begin{figure}
  \centering
\subfloat{\includegraphics[width=0.48\textwidth]{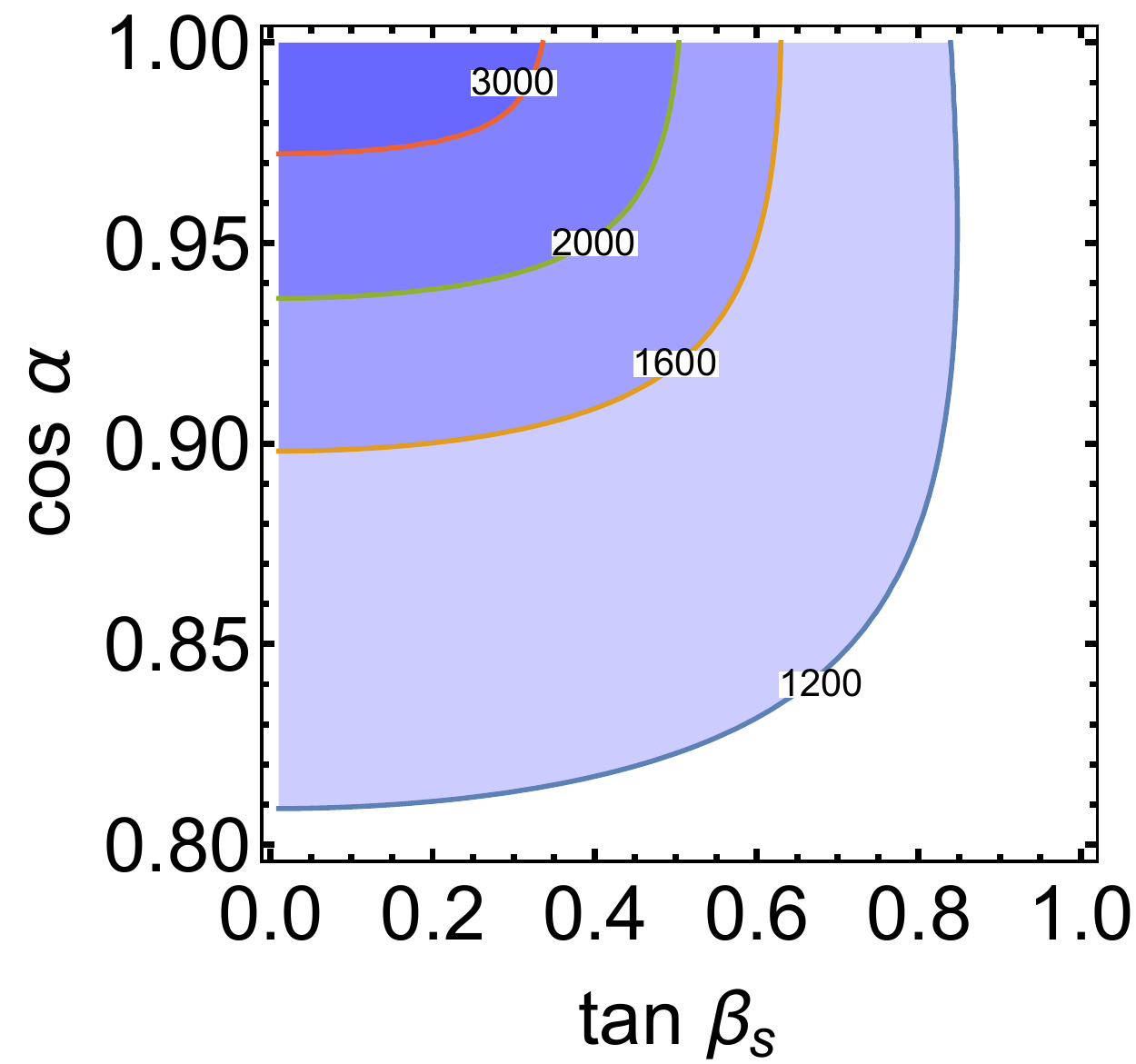}}
\subfloat{\includegraphics[width=0.48\textwidth]{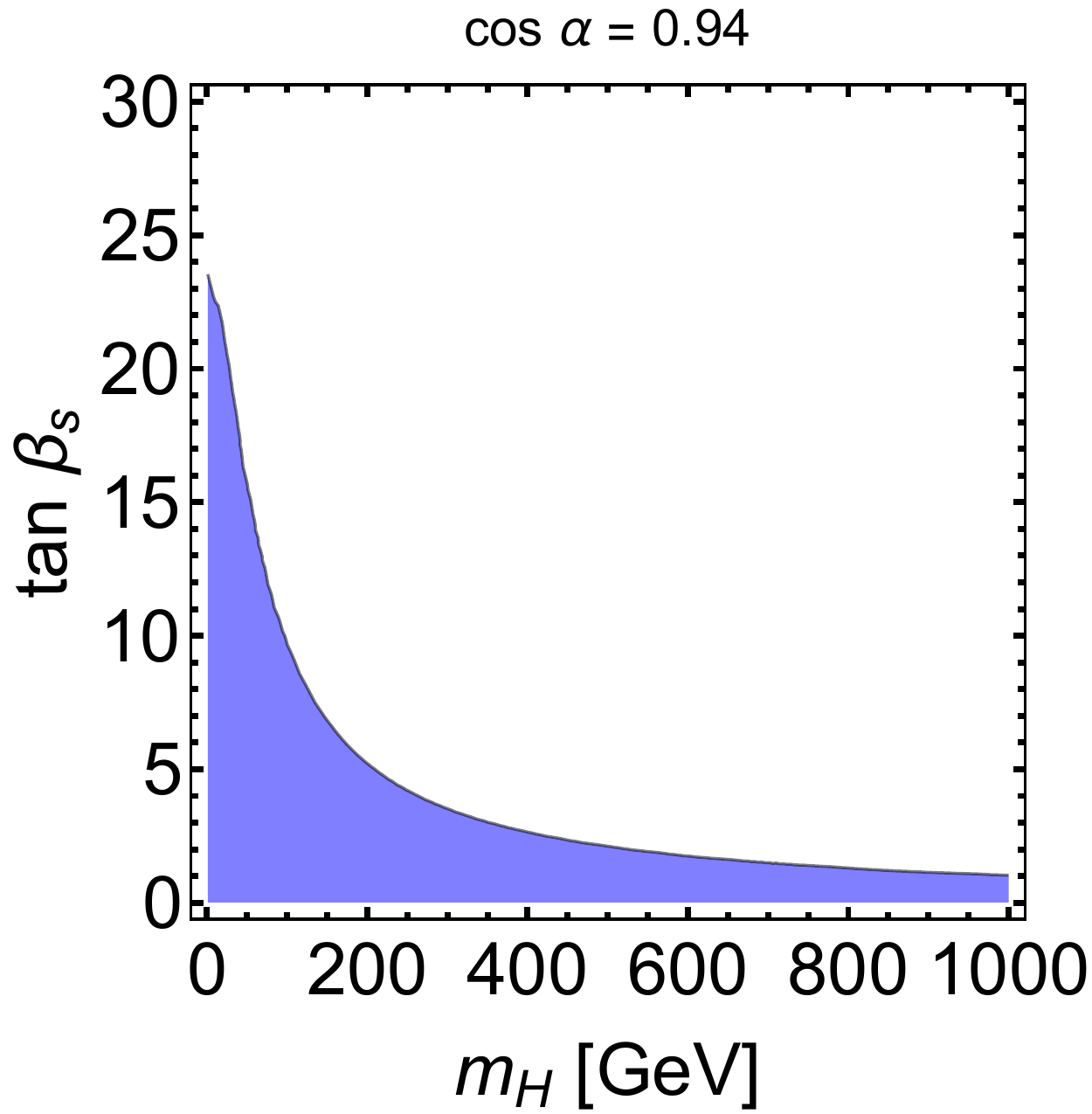}}
 \caption{Perturbativity bounds on the real singlet model with a spontaneously broken $Z_2$ symmetry. (Left:) The contours are labeled with the maximum value of $m_H /$GeV that is viable at that point. Going from lighter to darker shading indicates that heavier new scalars are allowed. (Right:) The shaded parameter space is allowed, and going to larger values of $\cos\alpha$ slightly increases the amount of viable parameter space, and darker shading indicates that heavier new scalars are allowed. }
  \label{fig:PUsingletZ2}
\end{figure}

We now turn our attention to the perturbativity bounds on the extended scalar sector theories. Using the method described above, typical bounds on the real singlet model with a spontaneously broken $Z_2$ symmetry are shown in Fig.~\ref{fig:PUsingletZ2}. In the left panel, the contours are labeled with the maximum value of $m_H /$GeV that is viable at that point. Darker shading indicates viable parameters space for a heavier new scalar. In the right panel, the shaded parameter space is allowed, and going to larger values of $\cos\alpha$ slightly increases the amount of viable parameter space. 
Considering only the high energy limit of ${\cal{H}}{\cal{H}}\rightarrow {\cal{H}}{\cal{H}}$, perturbative unitarity
requires,
\begin{equation}
m_H^2 < {16\pi v^2\over \tan^2\beta_s}\, ,
\end{equation}
which explains the general features of the RHS of Fig ~\ref{fig:PUsingletZ2}.
It is fair to say there is plenty of viable parameter space in this model.

\begin{figure}
  \centering
\includegraphics[width=0.48\textwidth]{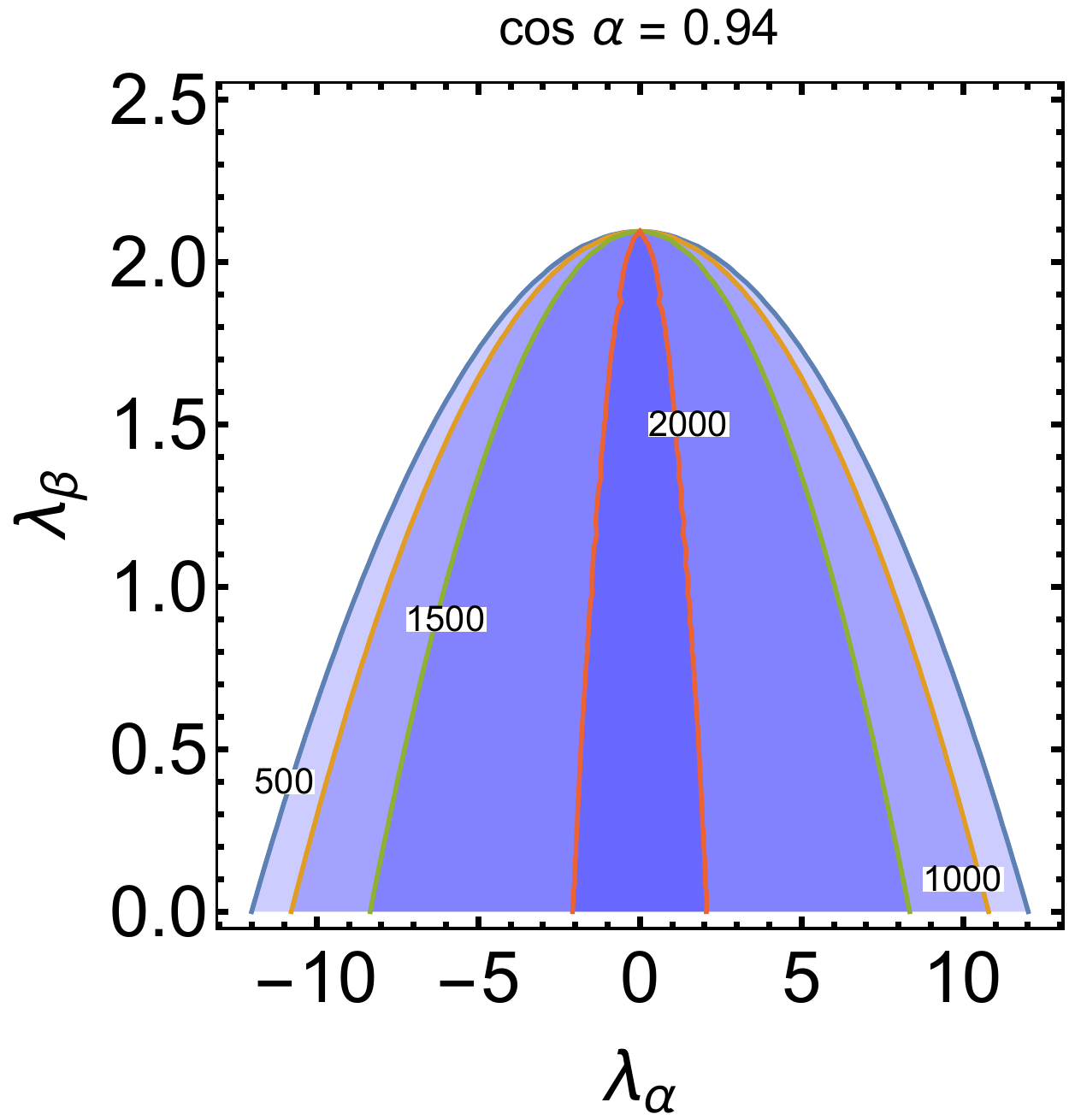}
 \caption{Perturbativity bounds on the real singlet model with an explicitly broken $Z_2$ symmetry. Just as in the left panel of Fig.~\ref{fig:PUsingletZ2}, the contours are labeled with the maximum value of $m_H /$GeV that is viable at that point, and going from lighter to darker shading indicates that heavier new scalars are allowed.}
  \label{fig:PUsinglet}
\end{figure}

The real singlet model with an explicitly broken $Z_2$ symmetry also has a fair amount of viable parameter space. This is illustrated in Fig.~\ref{fig:PUsinglet}. Just as in the left panel of Fig.~\ref{fig:PUsingletZ2}, the contours are labeled with the maximum value of $m_H$ that is viable at that point. From Fig.~\ref{fig:PUsinglet} we see that $\lambda_{\alpha}$, which enters into the Wilson coefficient $c_6$, is essentially a free parameter. Some of the viable parameter space in Fig.~\ref{fig:PUsinglet} will be ruled out by requiring the potential to be bound from below, $\lambda_{\alpha} > - 2 \sqrt{\lambda_{\beta} \lambda}$, but this does not significantly affect our conclusion.

\begin{figure}
  \centering
\subfloat{\includegraphics[width=0.52\textwidth]{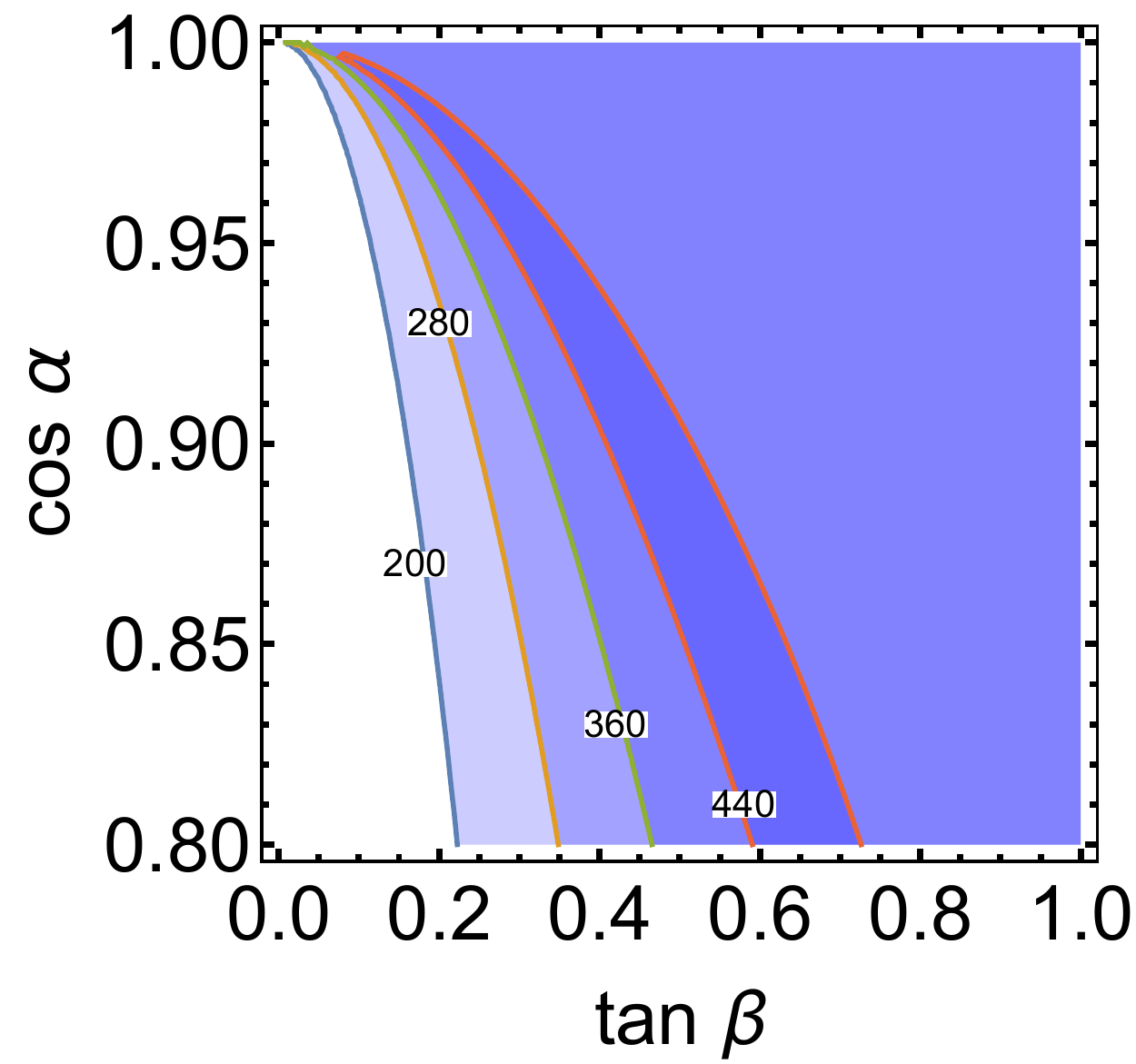}}
\subfloat{\includegraphics[width=0.48\textwidth]{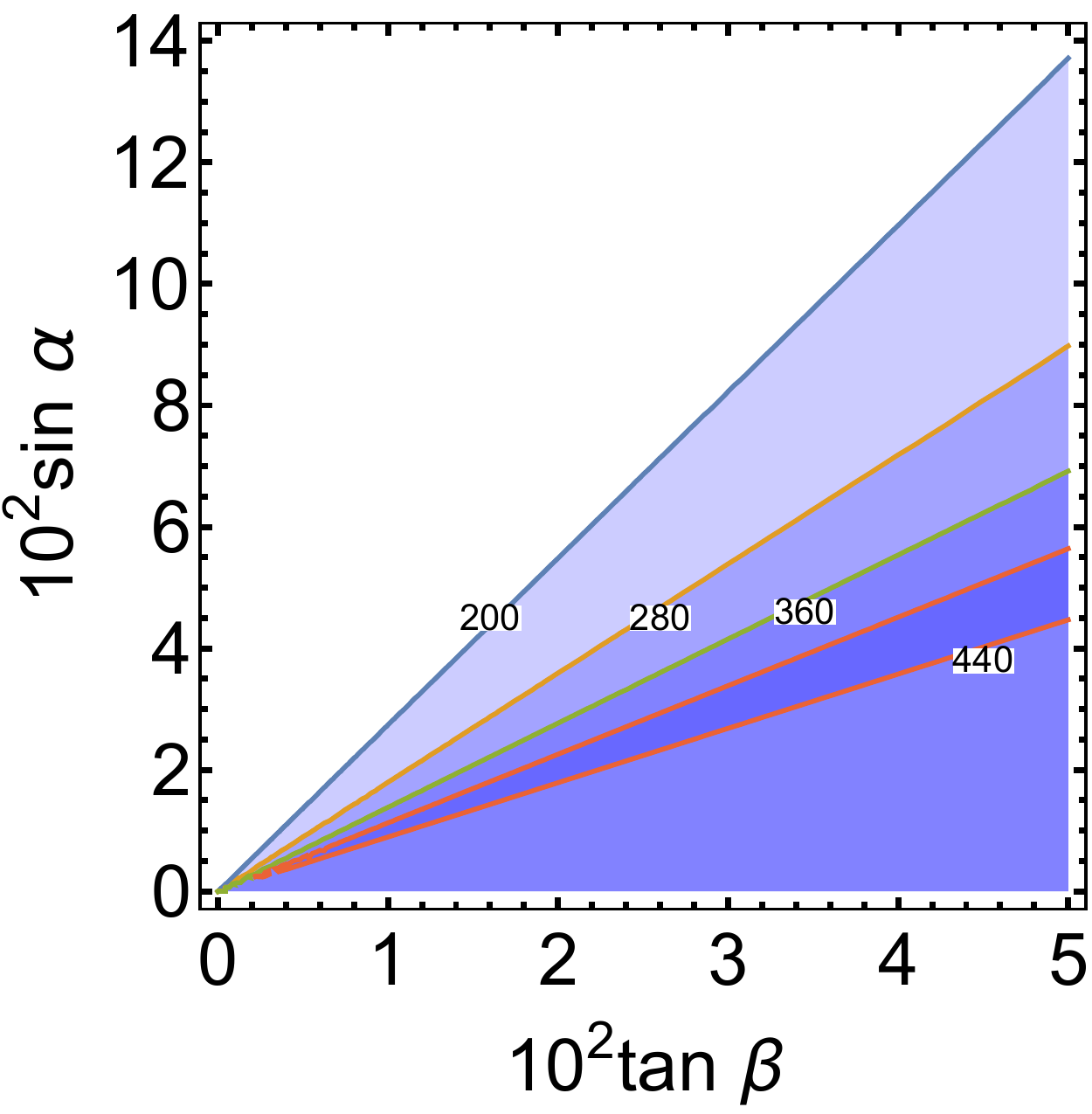}}
 \caption{Perturbativity bounds on the real triplet model. The contours are labeled with the maximum value of $m_H /$GeV that is viable at that point, with darker shading indicating heavier new scalars are allowed. For simplicity we have set the masses of all the heavy Higgs bosons to be equal. In the absence relatively light new scalars, combining the perturbativity bounds with the constraint on $\tan\beta$ from the $\rho$ parameter forces $\cos\alpha$ to be much closer to one than experimental measurements would otherwise require.}
  \label{fig:trip}
\end{figure}

\begin{figure}
  \centering
\subfloat{\includegraphics[width=0.52\textwidth]{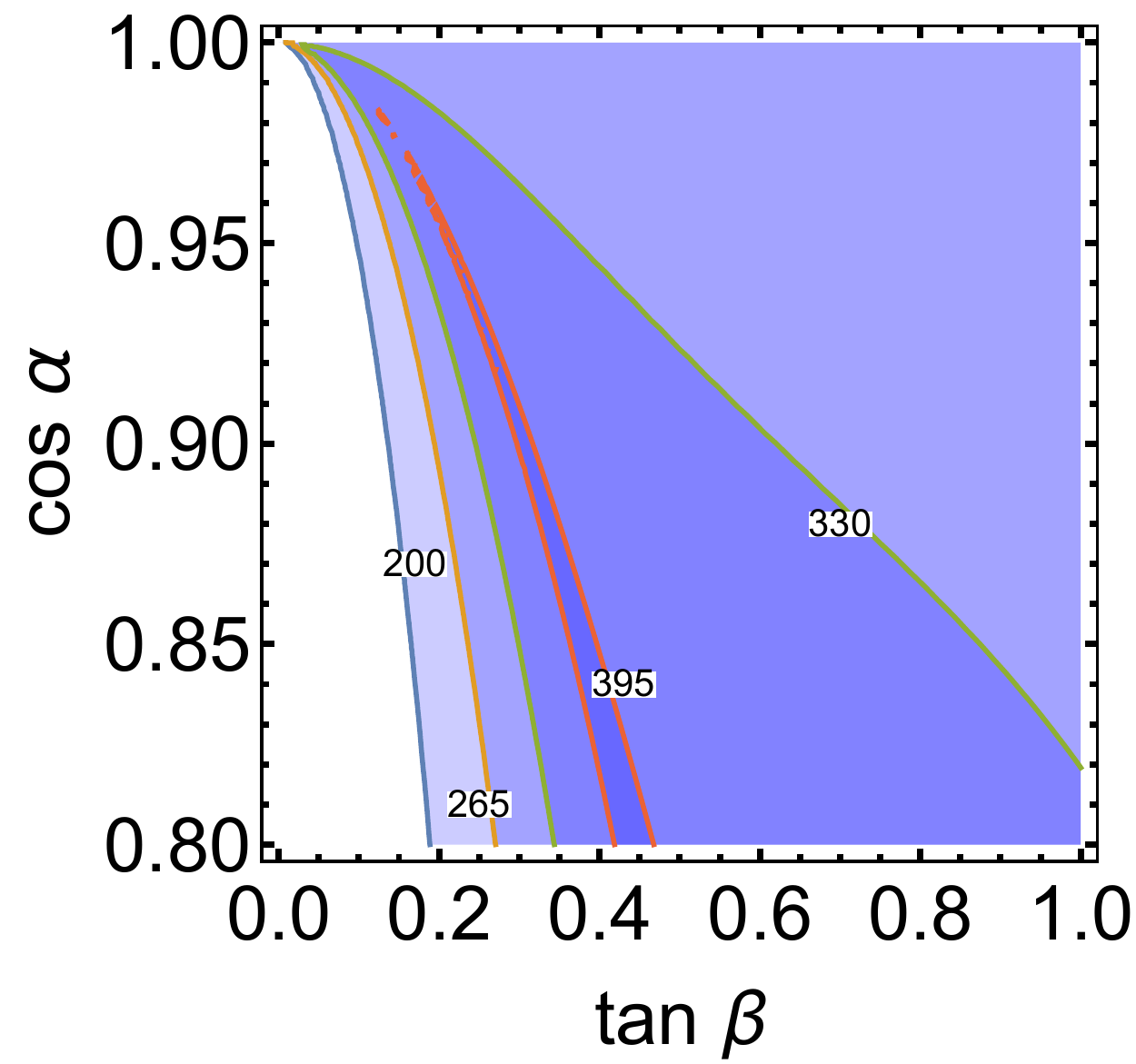}}
\subfloat{\includegraphics[width=0.48\textwidth]{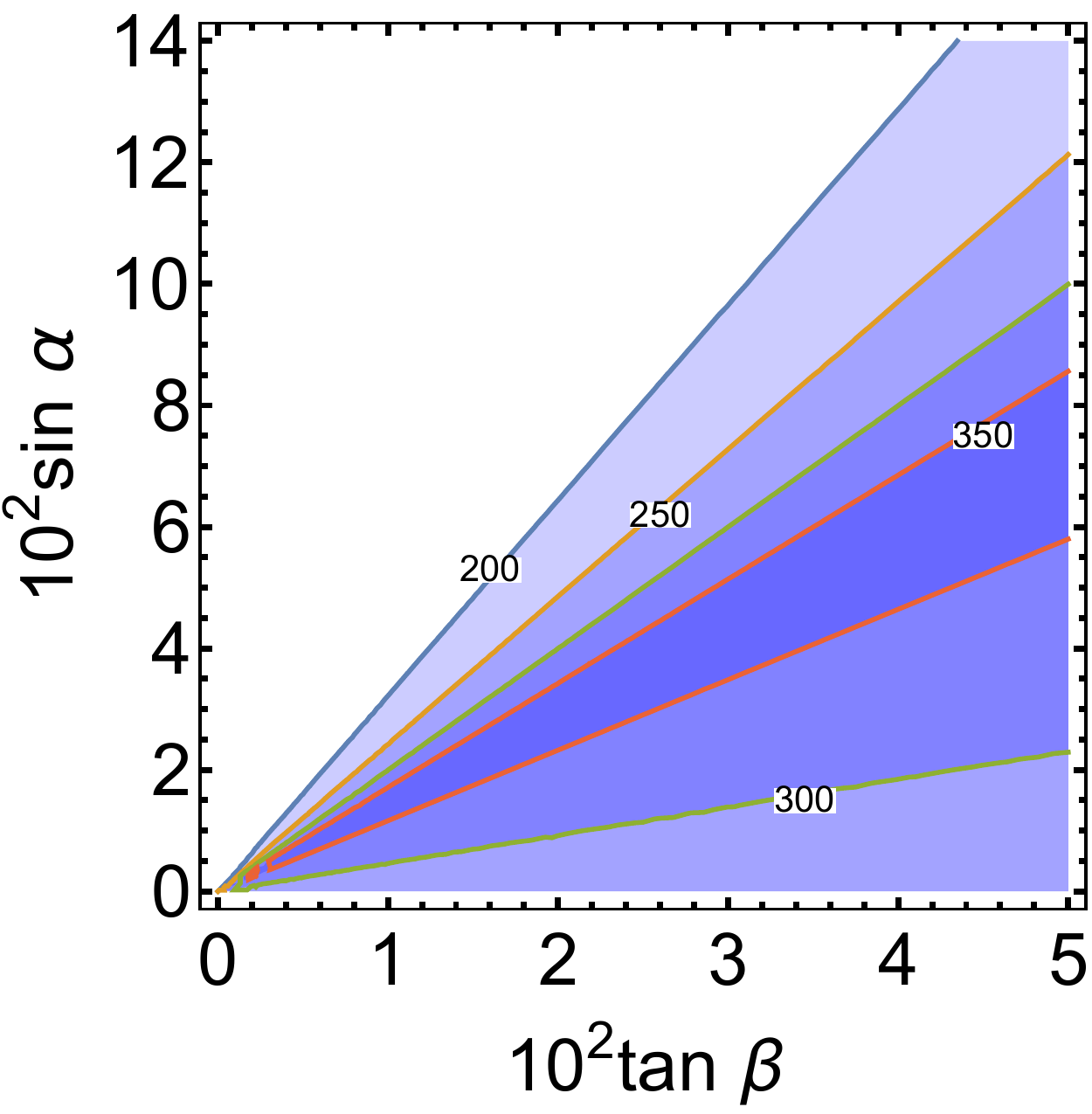}}
 \caption{The same as Fig.~\ref{fig:trip}, but for the complex triplet model.}
  \label{fig:tripC}
\end{figure}

\begin{figure}
  \centering
\subfloat{\includegraphics[width=0.52\textwidth]{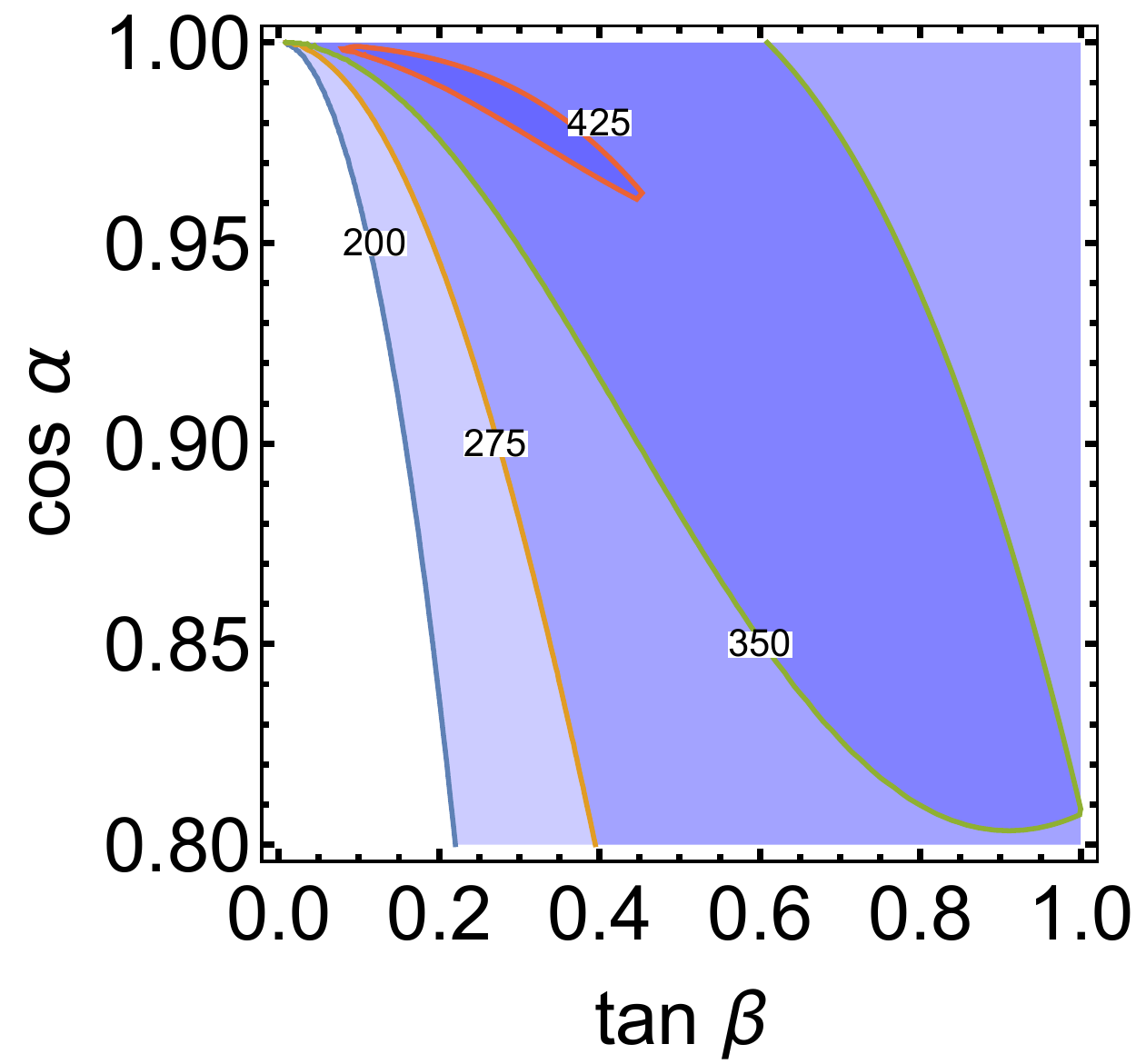}}
\subfloat{\includegraphics[width=0.48\textwidth]{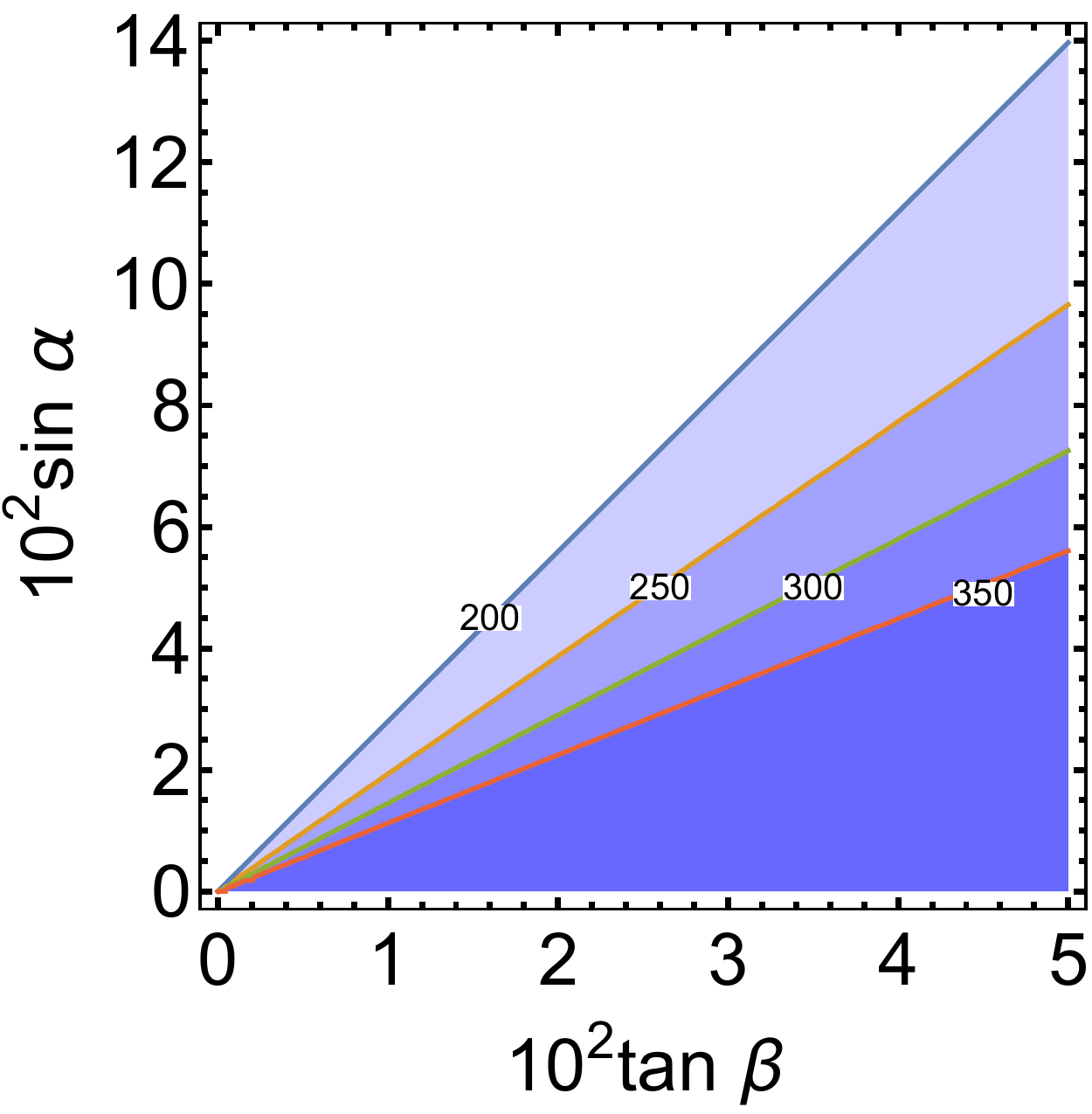}}
 \caption{The same as Fig.~\ref{fig:trip}, but for the quartet$_1$ model.}
  \label{fig:Q1}
\end{figure}

\begin{figure}
  \centering
\subfloat{\includegraphics[width=0.52\textwidth]{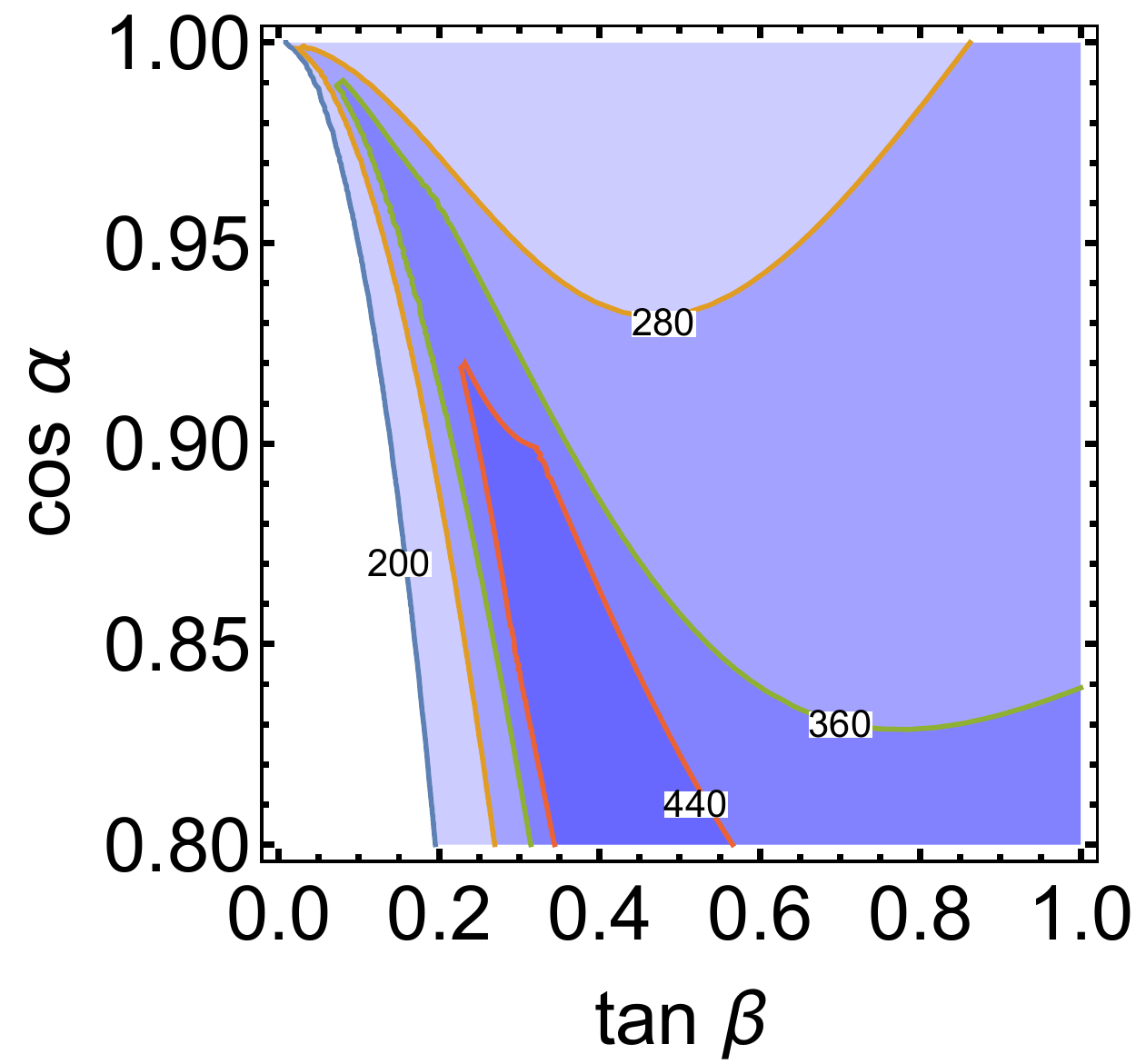}}
\subfloat{\includegraphics[width=0.48\textwidth]{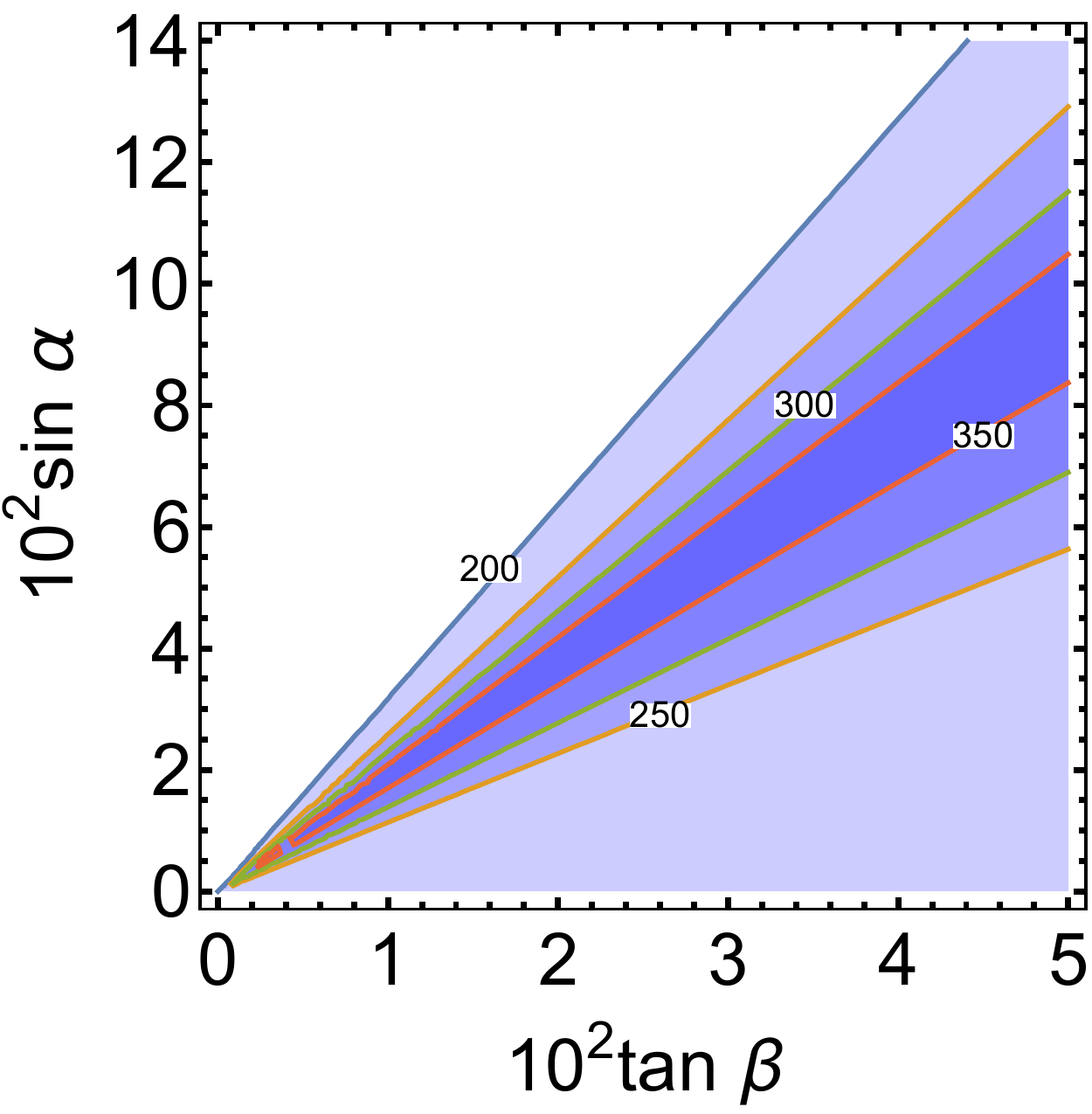}}
 \caption{The same as Fig.~\ref{fig:trip}, but for the quartet$_3$ model.} 
  \label{fig:Q3}
\end{figure}

The triplet and quartet models exhibit qualitatively similar behavior as far as perturbativity is concerned. The bounds on the real triplet model, with the simplifying assumption that the masses of the heavy Higgs bosons are equal, are shown in Fig.~\ref{fig:trip}. See Fig.~\ref{fig:tripC}, Fig.~\ref{fig:Q1}, and Fig.~\ref{fig:Q3} for the complex triplet, quartet$_1$, and quartet$_3$, respectively. We also assume, for simplicity, the masses of the heavy Higgs bosons are all equal in the complex triplet and quartet$_3$ models. On the other hand, for the quartet$_1$ model we take $\overline{m}_{H^+} = \sqrt{2} m_A$, (see Eq.~\eqref{eq:Cmasses})
and set the masses of the all the non-singly charged, heavy Higgs bosons to be equal. Furthermore, for the quartet$_1$ model we neglect the eigenvalues from the partial-wave matrices whose initial states had a net electric charge of either zero or one. These are $18 \times 18$ and $15 \times 15$ matrices, respectively, and thus are difficult to diagonalize. Explicitly, as an example, the singly charged initial scattering states in the Quartet$_1$ model are 
\begin{align}
i = &\{H^{++} H_2^-,\, H_1^+ {\cal{H}},\, H_1^+ A,\, H_2^+ {\cal{H}},\, H_2^+ A,\, H^{++} H_1^-,\, w^+ h, \\
& w^+ z,\, H^{++} w^-,\, H_1^+ h,\, H_1^+ z,\, H_2^+ h,\, H_2^+ z,\, w^+ {\cal{H}},\, w^+ A\} . \nn
\end{align}
Similarly, for the Quartet$_3$ model, we did not consider the eigenvalues from the partial-wave matrix whose initial states had a net electric charge of zero, which is a $16 \times 16$ matrix.

There are two main points we learn from Figs.~\ref{fig:trip},~\ref{fig:tripC},~\ref{fig:Q1}, and~\ref{fig:Q3}. Firstly, unless the `heavy' Higgs bosons are actually somewhat light, $m_H < 200$~GeV, combining the perturbativity bounds with the constraint on $\tan\beta$ coming from the $\rho$ parameter forces $\cos\alpha$ to be much closer to one than experimental measurements would otherwise require. Secondly, for given values of $\alpha$ and $\beta$ there are upper limits on the masses of the heavy Higgs bosons since no other parameters enter into the quartic couplings. We can use the first point to investigate the second point in more detail.

In the limit $\alpha \approx \beta \approx 0$, which is suggested by Figs.~\ref{fig:trip},~\ref{fig:tripC},~\ref{fig:Q1}, and~\ref{fig:Q3} as the only perturbative region consistent with the $\rho$ parameter, the expressions for the partial wave amplitudes simplify. This allows us to derive fairly simple analytic upper bounds on the masses of the heavy Higgs bosons or on the splittings between different masses in a multiplet. The  bounds for the real triplet model with $\alpha \approx \beta \approx 0$ are 
\begin{align}
469~\text{GeV} &> m_H , \\
\biggl({\beta\over 0.03}\biggr)^2\left(11.7~\text{GeV}\right)^2 &> \left| m_{H^+}^2 - m_H^2\right| . \nn
\end{align}
For numerical purposes we take $\beta$ to be at its upper limit, $\beta = 0.03$. We also neglected the mass of the 125 GeV Higgs boson in this analysis, which is justified \textit{a posteriori} since both the upper limit on $m_H^2$ and the mass splitting squared divided by $\beta^2$ are quite a bit larger than $m_h^2$. Comparable bounds are found in the other models.
\section{Double Higgs Production}
\label{sec:2Higgs}
\subsection{Formalism}
Double Higgs boson production in gluon fusion 
has been computed in Refs.~\cite{Glover:1987nx, Plehn:1996wb}. There have been many studies of double Higgs production using the both EFT approach as well as explicit models~\cite{Goertz:2014qta, Azatov:2015oxa, Cao:2016zob,Chen:2014ask, Dawson:2015haa, Cacciapaglia:2017gzh,DiLuzio:2017tfn,DiVita:2017eyz}.  
The SM rate can be found in Refs.~\cite{deFlorian:2016uhr, Borowka:2016ypz}.
The rate is dominated
by top quark loops, and for simplicity we neglect the $b-$loops. The SM rate is well below the current experimental limits from ATLAS and CMS~\cite{Khachatryan:2015yea, Aad:2015xja, CMS:2016zxv, Khachatryan:2016sey}.

Consider a theory with neutral scalars, $h$ and ${\cal{H}}$, and
 non-standard scalar cubic couplings and top Yukawa couplings parameterized as follows,
\begin{equation}
\mathcal{L} \supset - \frac{1}{6} \lambda_{hhh} v h^3 - \frac{1}{2} \lambda_{hhH} v h^2 {\cal{H}} - 
y_{ht} \frac{m_t}{v} \bar{t} t h - y_{Ht} \frac{m_t}{v} \bar{t} t {\cal{H}} .
\end{equation}
Expressions for $\lambda_{hhh}$ and $\lambda_{hhH}$ in the extended scalar models are given in Appendix  \ref{sec:modD}.
In the models we consider $y_{ht} = \cos\alpha$ and $y_{Ht} = \sin\alpha$. 

The partonic cross section for double Higgs production is~\cite{Plehn:1996wb}
\begin{equation}
\frac{d \hat{\sigma}}{d t} = \frac{G_F^2 \alpha_s^2}{\left(16 \pi\right)^3} \left( \left| C_{\triangle} 
F_{\triangle}(s) + C_{\Box} F_{\Box}(s,t)\right|^2 + \left| C_{\Box} 
G_{\Box}(s,t)\right|^2\right) ,
\end{equation}
where we have included a factor of $\tfrac{1}{2}$ for identical particles. The coefficients are given by
\begin{equation}
C_{\triangle} = \sum_{H_i=h,{\cal{H}}} \lambda_{hhH_i} \frac{v^2}{s - m_{H_i}^2 + i m_{H_i} \Gamma_{H_i}} y_{H_i t}  , \quad C_{\Box} = y_{ht}^2 .
\end{equation}
The form factors simplify considerably in the limit $m_t \to \infty$ (see~\cite{Plehn:1996wb} for the full expressions),
\begin{equation}
F_{\triangle} \rightarrow \frac{2}{3} + \mathcal{O}\left(\frac{s}{m_t^2}\right) , \quad F_{\Box} 
\rightarrow - \frac{2}{3} + \mathcal{O}\left(\frac{s}{m_t^2}\right) , 
\quad G_{\Box} \rightarrow \mathcal{O}\left(\frac{s}{m_t^2}\right) .
\end{equation}
In the SMEFT, considering only the top quark, the coefficients appearing in the cross section for double Higgs production are
\begin{align}
\label{eq:2Heft}
C_{\triangle} &\approx \frac{3 m_h^2}{s - m_h^2 + i m_h \Gamma_h} \left(1 + c_6 - 2 c_H - c_t\right) - \left(c_H + 3 c_t\right) , \\
\quad C_{\Box} &\approx 1 - c_H - 2 c_t , \nn
\end{align}
where we have expanded to linear order in the Wilson coefficients. The second term on the right-hand side of the first line 
of Eq.~\eqref{eq:2Heft} comes from the contact interaction $\bar{t} t h^2$. Unlike the amplitude, the cross section is not expanded in the Wilson coefficients. This is ensures a positive definite cross section. 

The hadronic level invariant mass distribution for double Higgs production is
\begin{equation}
\frac{d\sigma\left(p p \to h h\right)}{dM_{hh}} = \frac{2 M_{hh}}{S}  f\!\!f_{gg}\left(\frac{M_{hh}^2}{S}, M_{hh}\right) \hat{\sigma}\left(g g \to h h\right),
\end{equation}
with $S$ being the square of the collider center-of-mass energy, $M_{hh}^2 = s$, and  $f\!\!f_{gg}$ is the gluon luminosity function
\begin{equation}
f\!\!f_{gg}\left(y, \mu_F\right) = \int_y^1 \! \frac{dx}{x}\, f_{g / p}\left(x, \mu_F\right) f_{g / p}\left(\frac{y}{x}, \mu_F\right) , 
\end{equation}
where $f_{i / p}$ is the proton parton distribution function (PDF) of species $i$, and $\mu_F$ is the factorization scale. The total cross section is obtained by integrating the invariant mass distribution over $M_{hh}$ from $2 m_h$ to $\sqrt{S}$. Unlike the invariant mass distribution, the transverse momentum distribution requires the differential partonic cross section,
\begin{equation}
\frac{d\sigma\left(p p \to h h\right)}{dp_T} = 2 p_T \int_{-y^{\star}}^{y^{\star}} \! dy \, \frac{2 s}{S}  f\!\!f_{gg}\left(\frac{s}{S}, \frac{\sqrt{s}}{2}\right) \frac{d\hat{\sigma}(g g \to h h)}{d t} . 
\end{equation}
The limit of integration for the rapidity is
\begin{equation}
y^{\star} = \frac{1}{2} \log\left(\frac{1 + \sqrt{1 - 4\left(m_h^2 + p_T^2\right) / S}}{1 - \sqrt{1 - 4\left(m_h^2 + p_T^2\right) / S}}\right) .
\end{equation}
Recall that $s = 4 \left(m_h^2 + p_T^2\right) \cosh^2(y)$ and $t = - p_T^2 - \left(m_h^2 + p_T^2\right) \exp\left(- 2 y\right)$.

In the case of a heavy scalar, the $p_T$ distribution is peaked near $m_H^2 = 4 \left(m_h^2 + p_T^2\right)$. One way to see this is by looking at the $p_T$ distribution in the narrow width approximation (NWA)
\begin{equation}
\frac{d\sigma\left(p p \to h h\right)_{NWA}}{dp_T} = \frac{\alpha_s^2 \lambda_{hhH}^2 y_{Ht}^2}{2048 \pi^2 S\Gamma_H}  f\!\!f_{gg}\left(\frac{m_H^2}{S}, \frac{m_H}{2}\right) \frac{p_T}{\sqrt{m_H^2 - 4 \left(m_h^2 + p_T^2\right)}} + \mathcal{O}\left(\Gamma_H^0\right). 
\end{equation}
The total cross section is finite in the narrow width approximation despite the pole in the $p_T$ distribution
\begin{equation}
\sigma\left(p p \to h h\right)_{NWA} = \frac{\alpha_s^2 \lambda_{hhH}^2 y_{Ht}^2}{8192 \pi^2 S} \frac{m_H}{\Gamma_H} \sqrt{1 - \frac{4 m_h^2}{m_H^2}} f\!\!f_{gg}\left(\frac{m_H^2}{S}, \frac{m_H}{2}\right) + \mathcal{O}\left(\Gamma_H^0\right).
\end{equation}
\subsection{Numerical Results}

In this section, we compare predictions for double Higgs production at the LHC in the singlet, triplet, and quartet models with
predictions from the dimension-6 SMEFT.  We choose input parameters for mixing and masses consistent with restrictions from perturbative unitarity and the $\rho$ parameter.  
We use CT12NLO~\cite{Owens:2012bv} PDFs with a scale choice, $\mu_R=\mu_F=\sqrt{s}$. We include the full top quark
mass dependence, and neglect the small contribution from the $b$ quark.  

\subsubsection{Singlet Model with spontaneously broken $Z_2$ symmetry}
In the SMEFT for the singlet model with a spontaneously broken $Z_2$ symmetry, only $c_H$  is non-zero, and we employ the large mass limit for the SMEFT results, $c_H\sim \tan^2\alpha$, in our plots. 
In this model, the $Z_2$ symmetry imposes $c_6=0$~\cite{Henning:2014wua, deBlas:2014mba, Gorbahn:2015gxa}. 

\begin{figure}
\begin{centering}
\includegraphics[width=0.48\textwidth,clip]{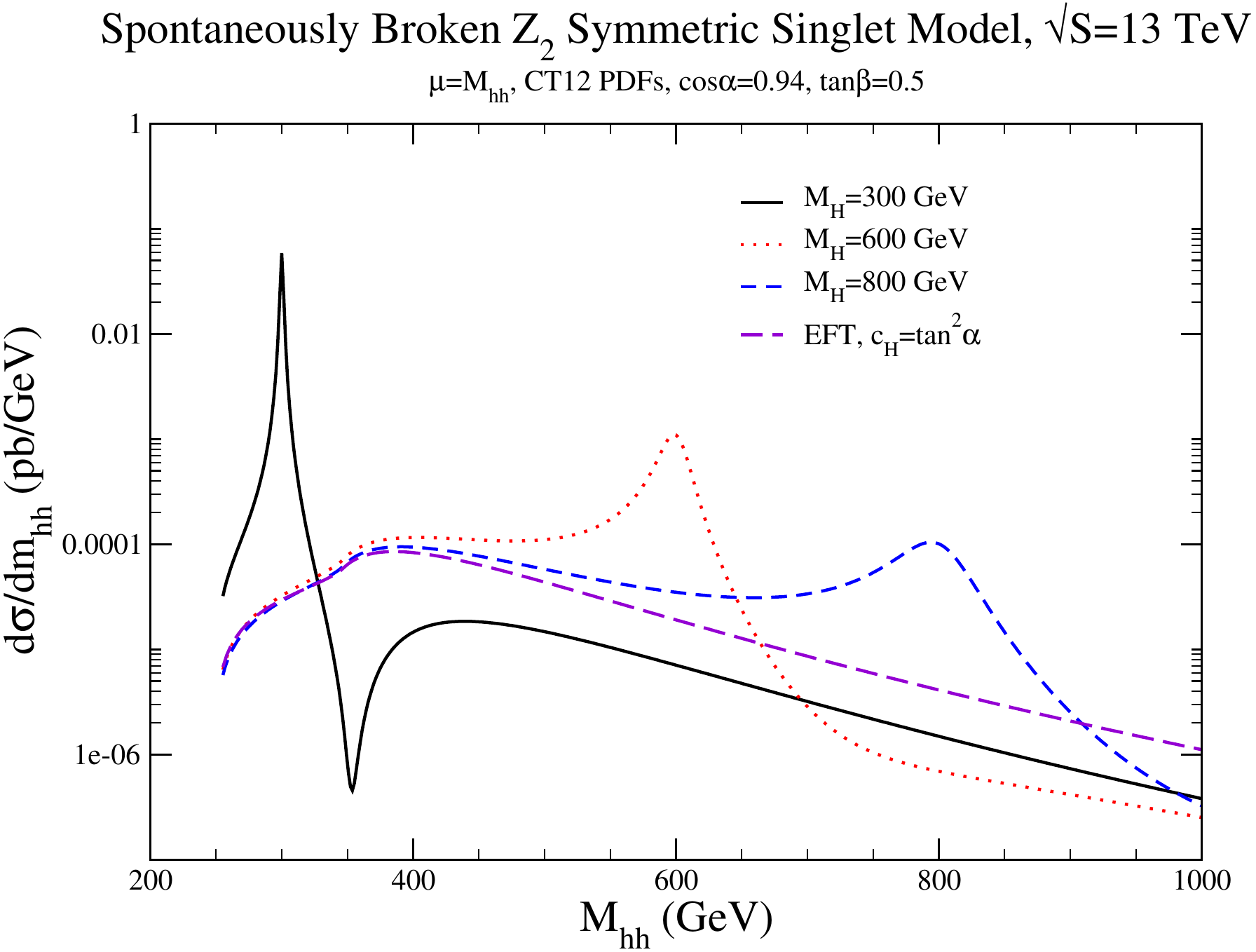} 
\includegraphics[width=0.48\textwidth,clip]{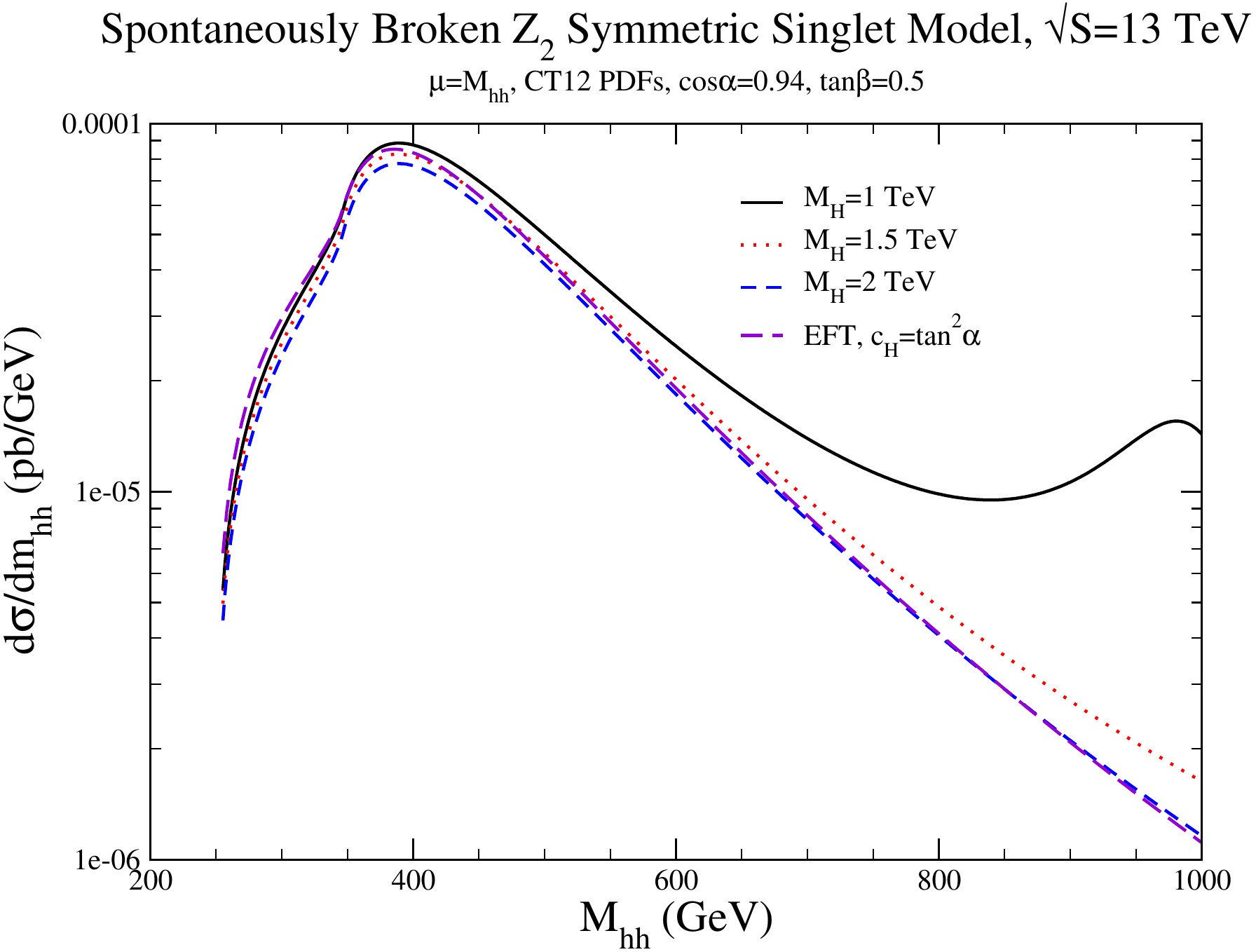}
\par\end{centering}
\caption{$d\sigma/dM_{hh}$ in the spontaneously broken $Z_2$  singlet model compared with the SMEFT predictions.  }
\label{fig:singz213}
\end{figure}

\begin{figure}
\begin{centering}
\includegraphics[width=0.48\textwidth,clip]{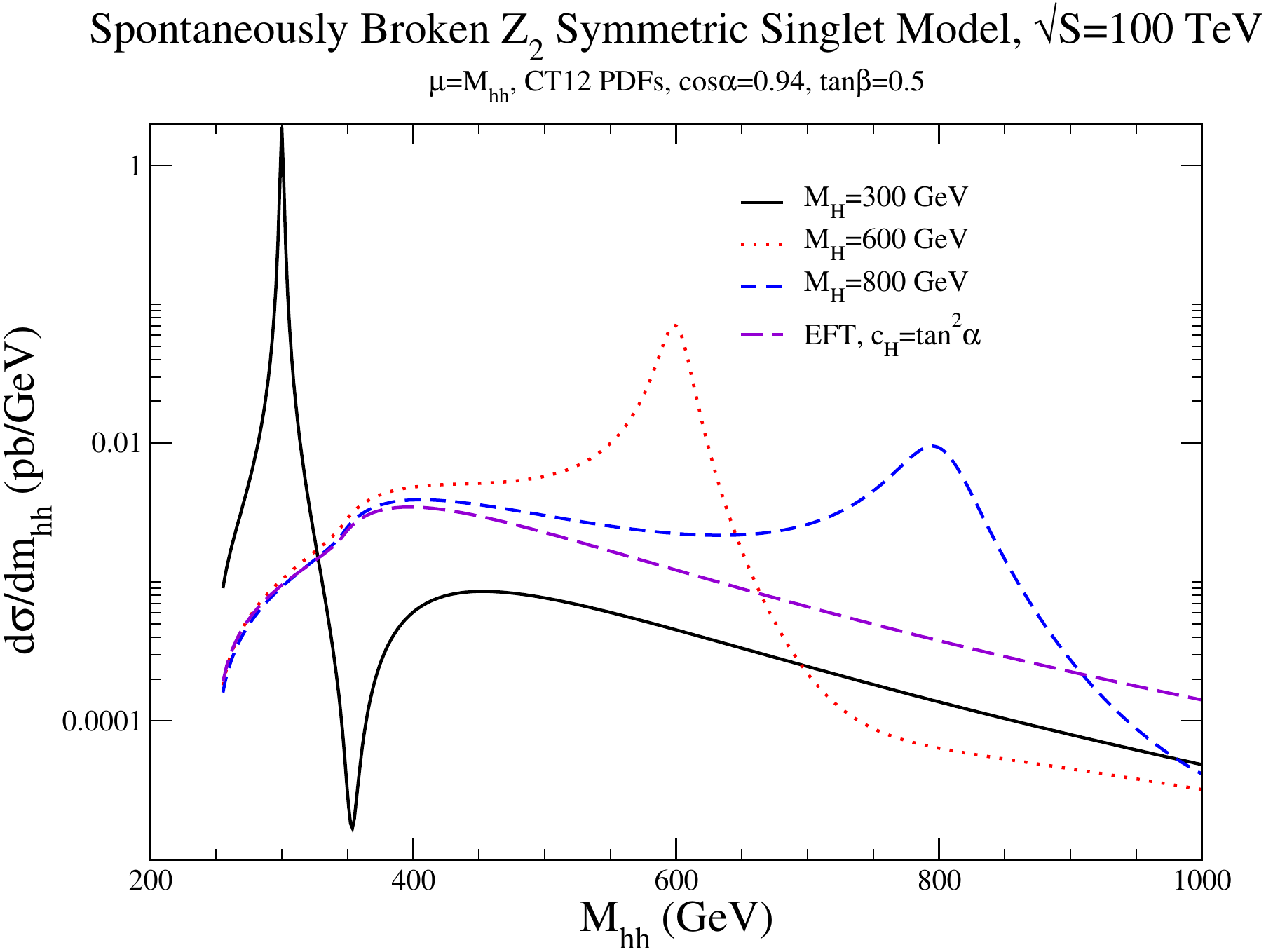} 
\includegraphics[width=0.48\textwidth,clip]{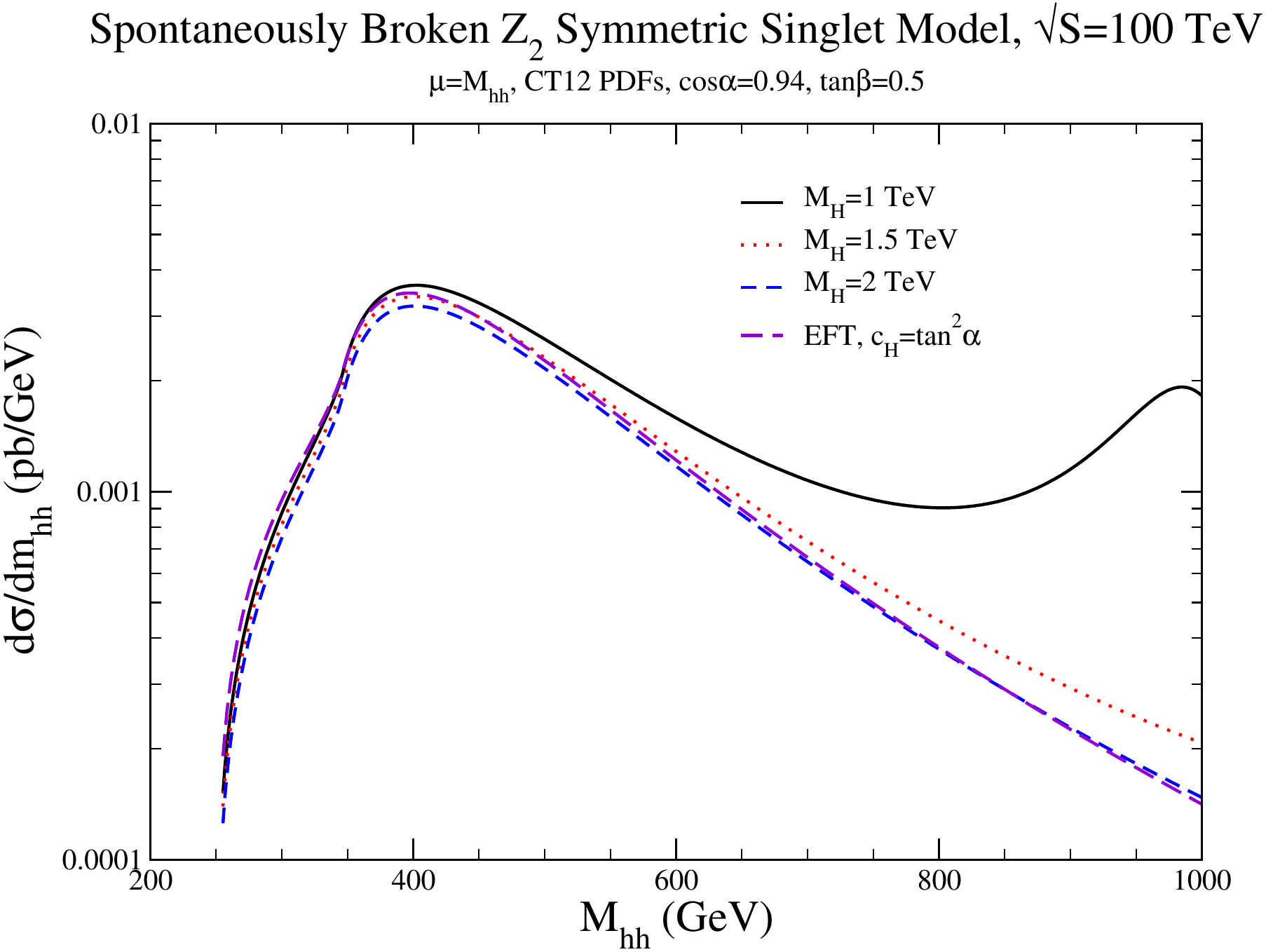}
\par\end{centering}
\caption{$d\sigma/dM_{hh}$ in the  spontaneously broken $Z_2$ singlet model compared with the SMEFT predictions. }
\label{fig:singz2100}
\end{figure}

The left-hand sides of Fig.~\ref{fig:singz213}
and Fig.~\ref{fig:singz2100} show the invariant mass distributions for the spontaneously broken $Z_2$ 
singlet model for heavy
Higgs masses of $m_H=300,\, 600$ and $800$~GeV for $\sqrt{S}=13$ and $100$~TeV.   The resonance peaks and
interference patterns are clearly observed.   The results in the SMEFT are also shown.  For $m_H=600$ and $m_H=800$~GeV,
the SMEFT is a good approximation to the invariant mass distribution below around $400$~GeV at both $\sqrt{S}=13$ and $100$~TeV.  Heavier masses are shown on the right-hand sides of Fig.~\ref{fig:singz213}
and Fig.~\ref{fig:singz2100}.  Below $M_{hh}\sim m_H/2$ the agreement between the singlet model results and the SMEFT limit
is excellent.  By the time $m_H$ reaches $2$~TeV, the SMEFT almost exactly reproduces the full model results.


\begin{figure}
\begin{centering}
\includegraphics[width=0.48\textwidth,clip]{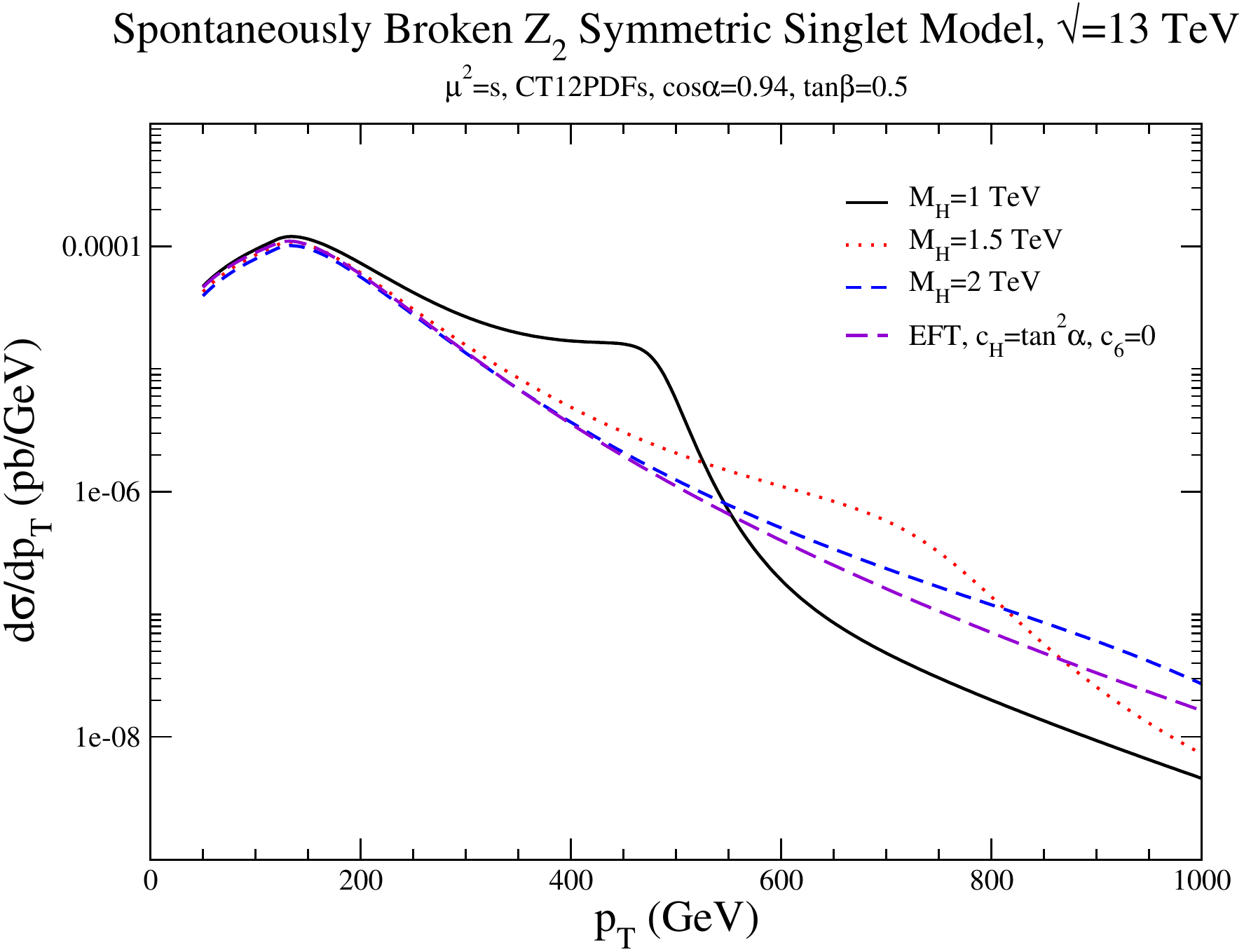} 
\includegraphics[width=0.48\textwidth,clip]{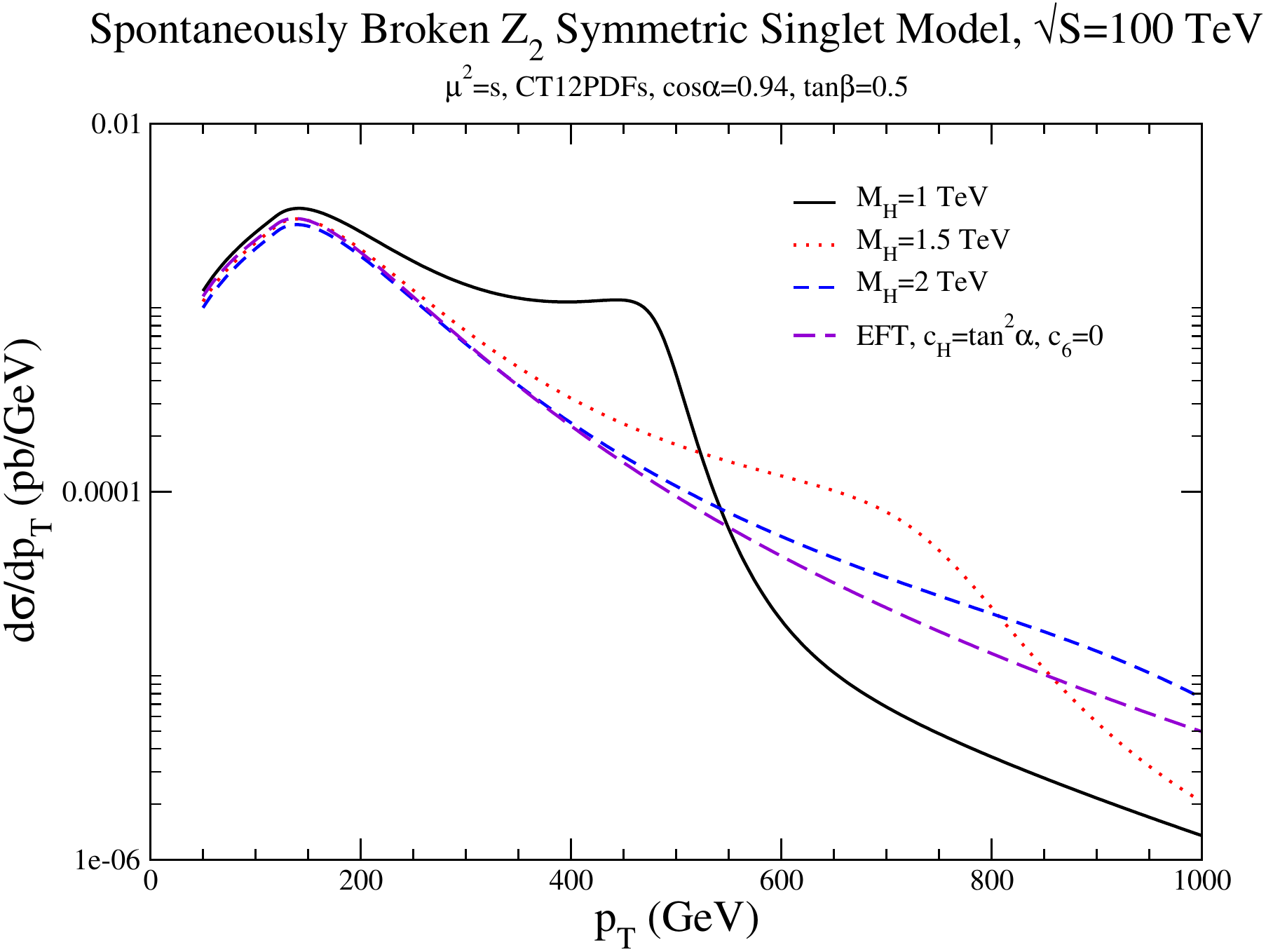}
\par\end{centering}
\caption{$d\sigma/dp_T$ in the spontaneously broken $Z_2$ singlet model compared with the SMEFT predictions. }
\label{fig:singz2pt}
\end{figure}

The $p_T$ distributions for the spontaneously broken $Z_2$ symmetric model are shown in Fig. \ref{fig:singz2pt}.  
For  $m_H\sim 1.5$~TeV,
the agreement below the resonance peak between the exact and SMEFT results is good below about $p_T\sim 400$~GeV,
while for $m_H=2$~TeV, the agreement is within a factor of $2$ even at $p_T=1$~TeV.

\subsubsection{Singlet Model with explicitly broken $Z_2$ Symmetry}
\begin{figure}
\begin{centering}
\includegraphics[width=0.48\textwidth,clip]{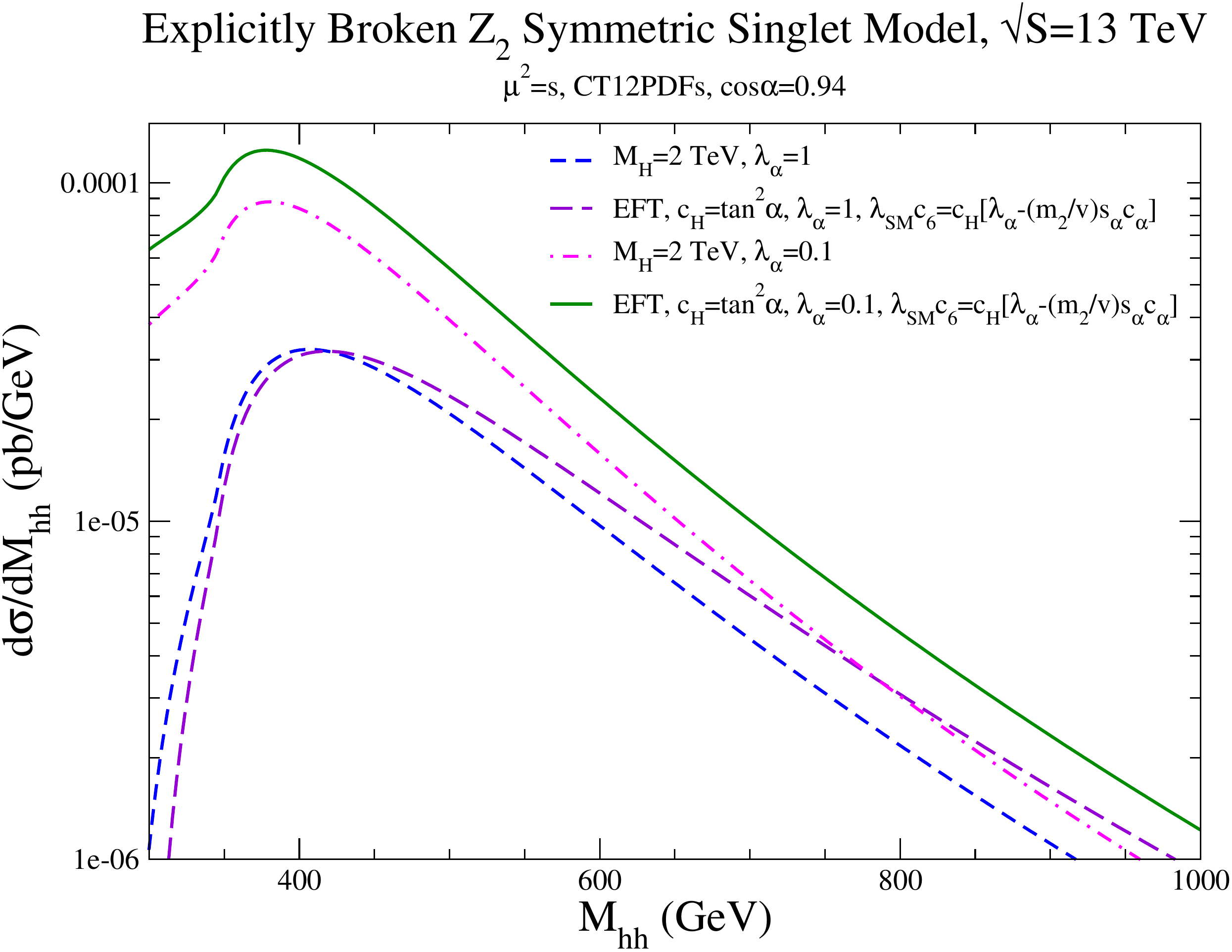} 
\par\end{centering}
\caption{$d\sigma/dM_{hh}$ in the  explicitly broken  $Z_2$ singlet model compared with the SMEFT predictions. }
\label{fig:noz2}
\end{figure}

The singlet model with explicit breaking of the $Z_2$ symmetry is described  by $6$ parameters that we fix to be
$v,\, m_h,\, m_H,\, \alpha,\, \lambda_\alpha$, and  $m_2$.  We take $\cos\alpha=0.94, ~\lambda_\alpha=0.1$ 
($\lambda_\alpha=1$), and $m_2=v$
for our numerical study. These parameters are chosen to obey all constraints from unitarity and the $\rho$ parameter.
In Fig.~\ref{fig:noz2}, we show the $M_{hh}$ in the explicitly broken $Z_2$ singlet model and compare
it with the SMEFT predictions.  The new feature of this model is that $c_6$ is no longer forced to be zero and
can be tuned by adjusting $\lambda_\alpha$.  We see fairly good agreement between the full theory and the SMEFT
for $m_H\sim 2$~TeV.

\subsubsection{Triplet Models}

The triplet model is highly restricted by the experimental limit on the $\rho$ parameter and when parameters are chosen
so as to be consistent with the $\rho$ parameter and perturbative unitarity, the mixing angle $\alpha$ is forced to be 
so small as to make the $gg\rightarrow hh$ cross section indistinguishable from the SM result.   This is a case where
the new physics is not probed by either single or double Higgs production.

\subsubsection{Quartet Models}
\begin{figure}
\begin{centering}
\includegraphics[width=0.48\textwidth,clip]{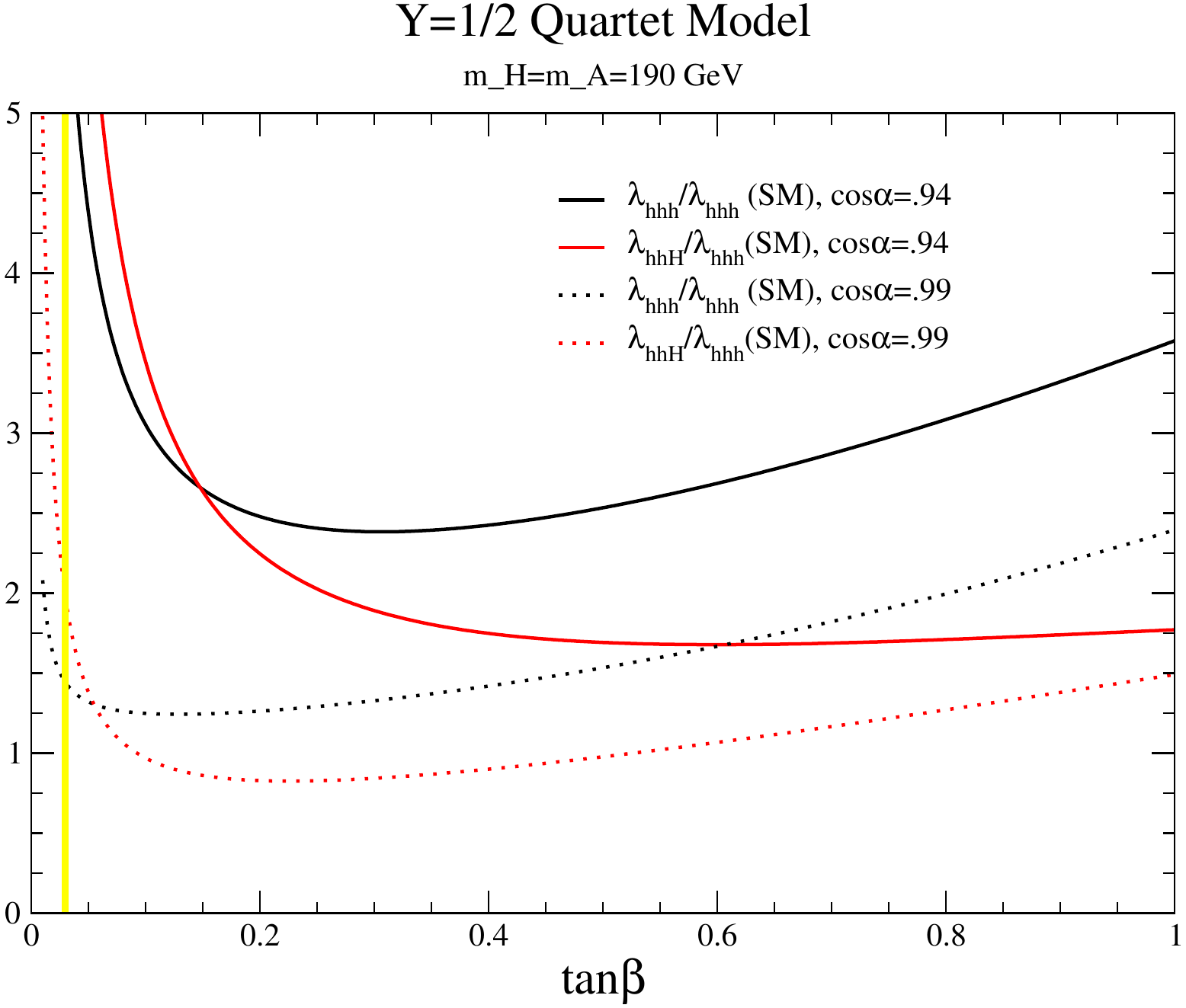}
\includegraphics[width=0.48\textwidth,clip]{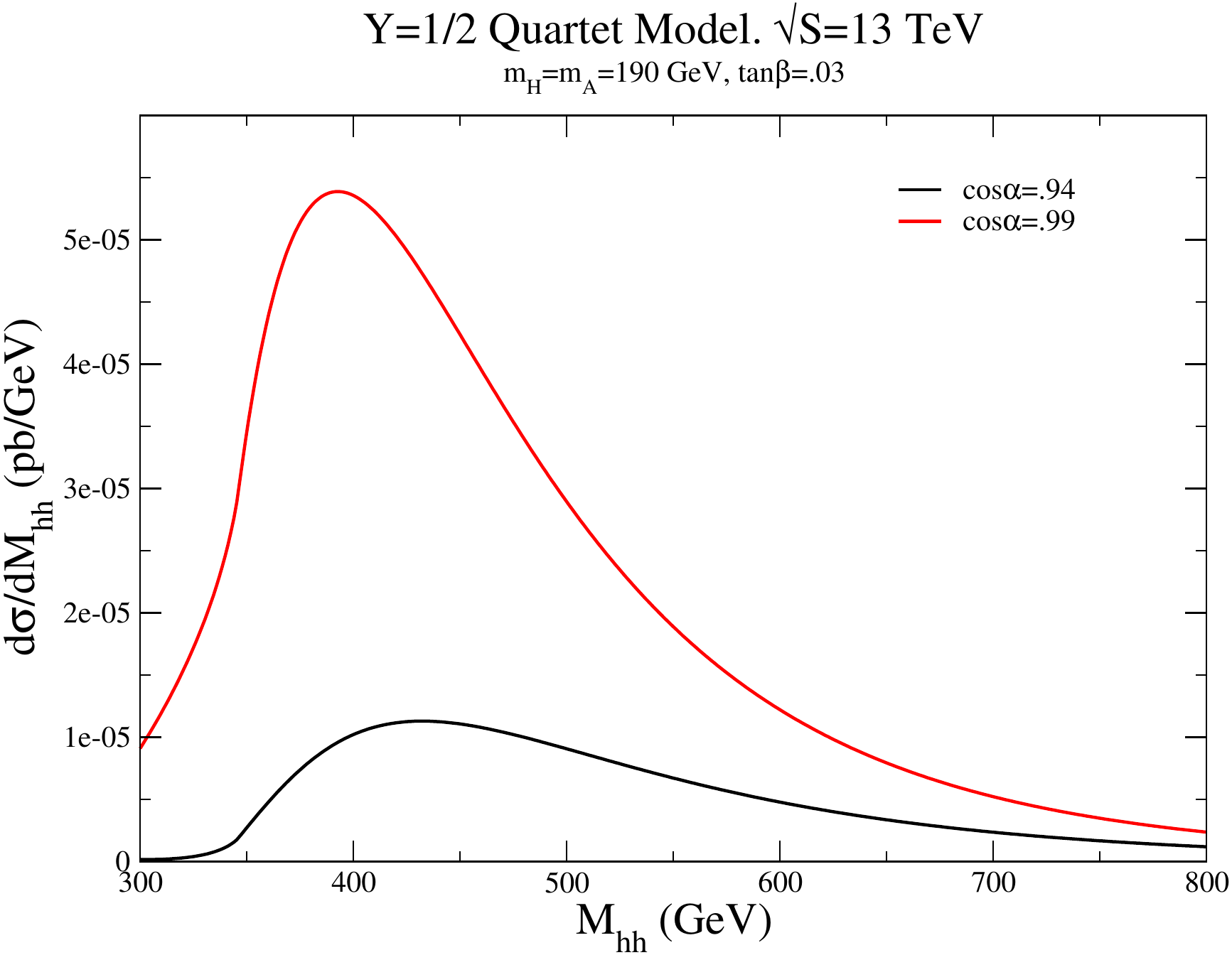} 
\par\end{centering}
\caption{Tri-linear Higgs couplings in the $Y={1\over 2}$ quartet model, normalized to the SM result (LHS).  The vertical
yellow line is the largest value of $\tan\beta=0.03$ allowed by perturbative unitarity for the chosen masses. The RHS shows
the invariant mass distribution for $2$ choices for the neutral mixing angle, $\alpha$.  }
\label{fig:quadfig}
\end{figure}

From the previous sections, we see that the limit on the $\rho$ parameter requires $\beta < 0.033$ for the quartet$_1$ model and $\beta < 0.010$ for the quartet$_3$ model.  For small $\tan\beta$ and small mixing $\alpha$,  
perturbative unitarity allows small regions of parameter space where the scalar masses are fine tuned. The allowed scalar masses
are electroweak scale, so the SMEFT is not applicable. 
In Fig~ \ref{fig:quadfig}, we compare the tri-linear Higgs coupling to the SM coupling for allowed parameters in the quartet$_1$ model.
For $\tan\beta\rightarrow 0$ and $\sin\alpha\rightarrow 0$, the SM is recovered, although $\cos\alpha=0.94$ gives 
significant deviations in the $hhh$ coupling from the SM result.
When the $hhh$ coupling is  non-SM like, the SM cancellation between the triangle and box contributions to $gg\rightarrow hh$ is spoiled,  and the results differ significantly from the SM.  This is clear in the $\cos\alpha=0.94$ curve on the RHS of Fig. \ref{fig:quadfig}.
%

\section{Conclusions}
\label{sec:con}

We have considered modifications of the SM with additional $SU(2)_L$ Higgs singlets, triplets, and quartets and computed their
contributions to SMEFT coefficients in the limits that the new scalars are heavy.  The coefficients show a characteristic pattern in the heavy mass limit, shown in Tab.~\ref{tab:dim6ops}.  A feature of the extended scalar models considered here is
that they generate only a subset of the possible SMEFT operators.  A fit to these operators from single Higgs production 
(Fig.~\ref{fig:cHcf}) shows that the data can not yet distinguish between the extended scalar sectors considered here,
although the sign of the $c_H$ is a generic signature of the UV scalar multiplet.

The parameters of the extended sectors are restricted by measurements of the $\rho$ parameter (see Tab.~\ref{tab:rho}) and perturbative unitarity.  In the triplet and quartet models, the $\rho$ parameter limits typically force $\tan\beta$ to be small, while in all models,
the requirement of perturbative unitarity puts an upper limit on  the heavy neutral scalar, $m_H$,  for a given value of $\tan \beta$.

There are regions of parameter space in the triplet model that are consistent with limits from the $\rho$ parameter, perturbative
unitarity, and single Higgs production.  In these models, the mixing between the two neutral Higgs
bosons is forced to be so small that both single
and double Higgs production look SM-like.  These models must be probed by searches for the charged scalars, which
are required by perturbative unitarity to be rather light.  The quartet models examined here have 
small regions of scalar masses simultaneously allowed by the $\rho$ parameter and perturbative unitarity limits.
In both the triplet and quartet models, however, the scalars are typically forced to be electroweak scale making the SMEFT not 
applicable.

The most interesting model considered here is the singlet model.  We have considered  models with a spontaneously
broken and an explicitly broken $Z_2$ symmetry.  Limits from perturbative unitarity require $\tan\beta_s <1$ in the spontaneously 
broken model, but allow
for a TeV scale neutral Higgs boson.  Comparisons of the invariant mass and $p_T$ distributions from double Higgs production
in the singlet models show that for $m_H\gsim 2$~TeV, the agreement between the exact calculations and the 
SMEFT results is excellent.
\begin{acknowledgments}
This work was supported by the United States Department of Energy under Grant Contract DE-SC0012704.
\end{acknowledgments}

\appendix
\section{EFT Details}
\label{sec:EFTd}

To help set the notation, consider the scalar potential of the SM
\begin{equation}
\label{eq:SMpot}
V_{SM} = - \mu^2 H^{\dagger} H + \lambda \left(H^{\dagger} H\right)^2 ,
\end{equation}
where $H^T = \left(w^+,\, (v + h + i z) / \sqrt{2}\right)$ with $w^{\pm}$ and $z$ being the Goldstone bosons, and $h$ is the physical Higgs scalar. The vev of the Higgs fields is set by minimizing the potential~\eqref{eq:SMpot}, $v = \mu / \sqrt{\lambda}$. The tree level mass for the Higgs boson is given by $m_h^2 = 2 \lambda v^2$. Finally, the equation of motion for the Higgs in the SM is~\cite{Buchmuller:1985jz,Jenkins:2013zja}
\begin{equation}
\label{eq:EoM}
D^2 H_k = \lambda v^2 H_k +2 \lambda \left(H^{\dagger} H\right) H_k - \bar{q}^j y_u^{\dagger} u \epsilon_{jk} + \bar{d} y_d q_k + \bar{e} y_e \ell_k \equiv EoM,
\end{equation}
where $j,\, k$ are $SU(2)$ indices, the flavor indices are implicit, and $\epsilon = \epsilon_{[jk]}$ with $\epsilon_{12} = +1$. 

Table~\ref{tab:ops} summarizes the possible dimension-6 operators involving $H$ and $D_{\mu}$, the relations amongst these operators, and how they are referred to in the literature~\cite{Grzadkowski:2010es, Khandker:2012zu, Contino:2013kra, Henning:2014wua}. It is straighforward to switch between operator bases, and convenient tables are given in Refs. \cite{Contino:2013kra,Falkowski:2015wza}.
\begin{table}
\centering
 \begin{tabular}{| c | c | c | c | c | c |}
 \hline
\textbf{Operator} & \textbf{Relation to}~\cite{Grzadkowski:2010es} & \multicolumn{4}{| c |}{\textbf{Operators, $O_i$}} \\ \hline
& References: &~\cite{Grzadkowski:2010es} &~\cite{Khandker:2012zu} &~\cite{Contino:2013kra} &~\cite{Henning:2014wua} \\
& Coefficients, ${\bar{C}}_i$: & $C_i$ & $a_i$ & $\frac{\bar{c}_i}{v^2}$ & $c_i$ \\ \hline
$\left|D^2 H\right|^2$ & $\left|EoM\right|^2$ & & & & $\mathcal{O}_D$ \\
$\left(H^{\dagger} H\right) \left(D_{\mu} H^{\dagger} D^{\mu} H\right)$ & $\frac{1}{2} Q_{H\Box} - \frac{1}{2} \left(H^{\dagger} H\right)\left(H^{\dagger} EoM + \text{ h.c.}\right)$ & & $O_2$ & & $\mathcal{O}_R$ \\
$ \left(\partial_{\mu} \left(H^{\dagger} H\right)\right)^2$ & $- Q_{H\Box}$ & & & $2 O_H$ & $2 \mathcal{O}_H$ \\
$\left|H^{\dagger} \overleftrightarrow{D}_{\mu} H\right|^2$ & $- Q_{H\Box} - 4 Q_{HD}$ & & & $2 O_T$ & $2 \mathcal{O}_T$ \\
$\left(H^{\dagger} H\right) \Box \left(H^{\dagger} H\right)$ & & $Q_{H\Box}$ & $2 O_1$ & & \\
$ \left| H^{\dagger} D_{\mu} H\right|^2$ & & $Q_{HD}$ & $O_T$ & & \\
$\left(H^{\dagger} H\right)^3$ & & $Q_H$ & \checkmark & $\frac{1}{\lambda} O_6$ & $\mathcal{O}_6$ \\ \hline
  \end{tabular}
  \caption{Summary of dimension-6 operators involving $H$ and $D_{\mu}$, including relations amongst the operators and the notation in the literature.  The effective Lagrangian in each basis is $L=\Sigma_i {{\bar C}_i}O_i$. }
  \label{tab:ops}
\end{table}

Redundant operators may appear in intermediate steps of calculations. An example of such an operator is $O_R = H^{\dagger} H \left(D_{\mu} H\right)^{\dagger} \left(D^{\mu} H\right)$ with coefficient $c_R / v^2$. To extract the coefficients of the $D^2 H^4$ operators  we follow the approach of~\cite{Khandker:2012zu}, which considers the following scattering processes evaluated at the matching scale
\begin{align}
\mathcal{M}\left(H_1(p) H_2(0) \to H_1(p) H_2(0)\right) &= c_R \frac{p^2}{v^2} , \\
\mathcal{M}\left(H_1(p) H_2(0) \to H_1(0) H_2(p)\right) &= \left(c_H - c_T\right) \frac{p^2}{v^2} , \nn \\
\mathcal{M}\left(H_1(p) H_2(-p) \to H_1(0) H_2(0)\right) &= \left(c_H + c_T\right) \frac{p^2}{v^2} . \nn
\end{align}
The subscripts in the above equations indicate the component of the Higgs doublet. The operator $\mathcal{O}_R$ can be removed with the following field definition~\cite{Giudice:2007fh,Low:2009di}
\begin{equation}
H \to H - \frac{c_R}{2 v^2} \left(H^{\dagger} H\right) H .
\end{equation}
This field definition leads to contributions to the non-redundant operators
\begin{equation}
c_H \to c_H - c_R , \quad c_f \to c_f + c_R / 2 .
\end{equation}

As previously mentioned, the kinetic energy for the Higgs boson, $h$, in Eq.~\eqref{eq:L} is not canonically normalized. This can be remedied by a simply rescaling~\cite{Grinstein:2007iv, Alonso:2013hga}
\begin{equation}
\label{eq:finrenorm}
h \to h / \sqrt{1 + c_H} .
\end{equation}
Alternatively, a field redefinition can be made to correctly normalize the  kinetic energy and eliminate the derivative interactions~\cite{Giudice:2007fh, Buchalla:2013rka}, Eq.~\eqref{eq:redef}. We stress that the two approaches yield equivalent results for physical observables, as expected. Using~\eqref{eq:redef} the Lagrangian now takes the form of Eq.~\eqref{eq:LeftCan}.
For example, both approaches lead to the following amplitude for Higgs-Higgs scattering 
\begin{align}
\mathcal{M}\left(h h \to h h\right) &= - \frac{3 m_h^2}{2 v^2} \left[2 + 12 c_6 + \frac{50}{3} c_H \right. \\
&\left.+ 3 \left(1 + 2 c_6 - 3 c_H\right) \left(\frac{m_h^2}{s - m_h^2} + \frac{m_h^2}{t - m_h^2} + \frac{m_h^2}{u - m_h^2}\right) \right] . \nn
\end{align}

\section{Scalar Model Details}
\label{sec:modD}

In this Appendix we give some additional details of the models considered in this work. The mixing angles analogous to $\alpha$ in~\eqref{eq:singmix} in the $CP$-odd and charged Higgs sectors are functions of $\beta$, and are denoted $\delta$ and $\gamma$, respectively.\footnote{There are three singly charged scalars in the quartet$_1$ model, and thus the diagonalization is more complicated in this case.}  Additionally, we will sometimes express the sine, cosine, or tangent of an angle $\theta$ as $s_{\theta}$, $c_{\theta}$, or $t_{\theta}$, respectively. Futhermore, we will sometimes use the following notation,
\begin{equation}
\lambda^{\prime}_i = \lambda_i v^2 / \rho, \quad m_i^{\prime} = m_i v / \rho, 
\end{equation}
with $\rho$ given in Tab.~\ref{tab:rho} for each model. 

\subsection{Real Singlet with Explicit $Z_2$ Breaking}
It is straightforward to compute the dimension-6 operators~\cite{Henning:2014wua} 
\begin{equation}
\label{eq:singH4D2}
c_6 \lambda_{SM} = \frac{m_1^2 v^2}{M^4} \left(\lambda_{\alpha} - \frac{m_1 m_2}{M^2}\right) , \quad c_H = \frac{m_1^2 v^2}{M^4}, \quad c_T = c_f = 0 .
\end{equation}
There are also shifts in the parameters of the renormalizable Lagrangian, for example, $\Delta \lambda = - m_1^2 / 2 M^2$. However these shifts are unphysical, and can simply be reabsorbed into the definition of the original parameters in the effective theory. We have checked by explicit computation that the matching is the same when starting either from the unbroken or broken phase of the full theory.

Now consider the masses and mixings in the full theory. The relations between the masses, mixing angle $\alpha$, and the Lagrangian parameters in the full theory (see~\eqref{eq:Vsing})  are
\begin{align}
4 \lambda v^2 &= m_h^2 + m_H^2 + \left(m_h^2 - m_H^2\right) \cos2\alpha , \\
2 M^2 &= m_H^2 + m_h^2 + \left(m_H^2 - m_h^2\right) \cos2\alpha , \nn \\
2 m_1 v &= \left(m_H^2 - m_h^2\right) \sin2\alpha . \nn
\end{align}

Lastly, the couplings in the full theory that are relevant for double Higgs production are
\begin{align}
\lambda_{hhh} v^2 / 3 &= m_h^2 \cos^3\alpha + 2 v \sin^2\alpha \left(\lambda_{\alpha} v \cos\alpha - m_2 \sin\alpha\right) , \\
2 \lambda_{hhH} v^2 / \sin\alpha &= \left(2 m_h^2 + m_H^2\right) \cos^2\alpha - 2 v \left(\lambda_{\alpha} v \left(1 + 3 \cos2\alpha\right) - 3 m_2 \sin2\alpha\right)  . \nn
\end{align}

\subsection{Real Singlet with Spontaneous $Z_2$ Breaking}
The only operator generated in this case is~\cite{Gorbahn:2015gxa}
\begin{equation}
c_H = \left(\frac{\lambda_{\alpha} v}{4 \lambda_{\beta} v_{\phi}}\right)^2 .
\end{equation}
In terms of mass eigenstates the quartic couplings in~\eqref{eq:VsingZ2} are
\begin{align}
4 \lambda v^2 &= m_h^2 + m_H^2 + \left(m_h^2 - m_H^2\right) \cos2\alpha , \\
4 \lambda_{\alpha} v_{\phi} v &= \left(m_H^2 - m_h^2\right) \sin2\alpha , \nn \\
16 \lambda_{\beta} v_{\phi}^2 &= m_H^2 + m_h^2 + \left(m_H^2 - m_h^2\right) \cos2\alpha , \nn
\end{align}
with $\alpha$ the same as in Eq.~\eqref{eq:singmix}. Interestingly, in the case of spontaneous $Z_2$ breaking, $c_H$ has the exact same form as~\eqref{eq:cHmassES} when written in terms of the physical masses and mixing angle. 

The couplings relevant for double Higgs production are somewhat simpler in this case
\begin{align}
\lambda_{hhh} &= \frac{3 m_h^2}{v^2} \left(\cos^3\alpha - \frac{v}{v_{\phi}} \sin3\alpha\right) , \\
\lambda_{hhH} &= \frac{2 m_h^2 + m_H^2}{2 v^2} \sin2\alpha \left(\cos\alpha + \frac{v}{v_{\phi}} \sin\alpha\right)  . \nn
\end{align}

\subsection{Real Triplet}
The coefficients of the dimension-6 operators are~\cite{Khandker:2012zu, Henning:2014wua},
\begin{equation}
\label{eq:reTripC}
c_H = - \frac{m_1^2 v^2}{2 M^4} , \quad c_T = - \frac{c_H}{2} , \quad c_6 \lambda_{SM} = - \frac{c_H}{2} \lambda_{\alpha} , \quad c_f = - \frac{c_H}{2} .
\end{equation}

Minimizing the potential ($V =$~\eqref{eq:Vrealtrip} $+ V_{\text{SM}}$) yields the constraints
\begin{align}
4 \mu^2 &= 4 \lambda v^2 c_{\beta}^2  - m_1 v s_{\beta} + \lambda_{\alpha} v^2 s_{\beta}^2, \\
2 M^2 v s_{\beta} &= m_1 v^2 c_{\beta}^2 - 2 \lambda_{\alpha} v^3 c_{\beta}^2 s_{\beta} - 2 \lambda_{\beta} v^3 s_{\beta}^3 . \nn
\end{align} 
The mixing angle in the charged sector is $\gamma = \beta$. The Lagrangian parameters can be traded for the masses of the particles and the mixing angles
\begin{align}
m_1 v &= 2 m_{H^+}^2 \sin\beta , \\
4 \lambda^{\prime} &= m_h^2 + m_H^2 + \left(m_h^2 - m_H^2\right) \cos2\alpha , \nn \\
\lambda_{\alpha} v^2 &= m_{H^+}^2 + \left(m_H^2 - m_h^2\right) \csc2\beta \sin2\alpha , \nn \\
4 \lambda_{\beta} v^2 &= 2 m_{H^+}^2 + \left(m_h^2 + m_H^2 - 2 m_{H^+}^2 + \left(m_H^2 - m_h^2\right) \cos2\alpha\right) \csc^2\beta . \nn
\end{align}
The cubic couplings relevant for double Higgs production are
\begin{align}
\lambda_{hhh} v^2 &= 3 m_h^2 \left(c_{\alpha}^3 c_{\beta}^{-1} - 2 s_{\alpha}^3 s_{\beta}^{-1}\right) + 6 m_{H^+}^2 s_{\alpha}^2 t_{\beta}^{-1} s_{\alpha + \beta} , \\
\lambda_{hhH} v^2 &= \left(2 m_h^2 + m_H^2\right) s_{\alpha} c_{\alpha} \left(c_{\alpha} c_{\beta}^{-1} + 2 s_{\alpha} s_{\beta}^{-1}\right) - m_{H^+}^2 s_{\alpha} t_{\beta}^{-1} \left(s_{\beta} + 3 s_{2\alpha + \beta}\right) . \nn
\end{align}

\subsection{Complex Triplet}
The coefficients of the dimension-6 operators are~\cite{Khandker:2012zu, deBlas:2014mba}
\begin{equation}
\quad c_H = - \frac{m_1^2 v^2}{2 M^4}, \quad c_T = c_H , \quad c_6 \lambda_{SM} = c_H \left(\frac{\lambda_{\alpha 2}}{2} - \lambda_{\alpha 1}\right) , \quad c_f = - c_H .
\end{equation}
Minimizing the potential ($V =$~\eqref{eq:Vcomptrip} $+ V_{\text{SM}}$) yields the constraints
\begin{align}
8 \mu^2 &= 8 \lambda v^2 c_{\beta}^2  + 4 \sqrt{2} m_1 v s_{\beta} + \left(2 \lambda_{\alpha 1} - \lambda_{\alpha 2}\right) v^2 s_{\beta}^2 , \\
- 2 M^2 s_{\beta} &= 2 \sqrt{2} m_1 v c_{\beta}^2 + \left( 2 \lambda_{\alpha 1} - \lambda_{\alpha 2}\right) v^2 c_{\beta}^2 s_{\beta} - 2 \left(\lambda_{\beta 1} + \lambda_{\beta 2}\right) v^2 s_{\beta}^3 . \nn
\end{align} 
The $CP$-odd and charged Higgs mixing angles are $\tan2\delta = (2 \sqrt{2} \sin2\beta) / (1 - 3 \cos2\beta)$ and $\gamma = - \beta$, respectively. In terms of the physical masses and $CP$-even mixing angle, the Lagrangian interaction parameters are
\begin{align}
- m_1^{\prime} &= \sqrt{2} m_A^2 \sin\beta , \\
4 \lambda v^2 \cos^2\beta &= m_h^2 + m_H^2 + \left(m_h^2 - m_H^2\right) \cos2\alpha , \nn \\
\lambda_{\alpha 1} v^2 &= 2 m_{H^+}^2 + \sqrt{2} \left(m_H^2 - m_h^2\right) \csc2\beta \sin2\alpha , \nn \\
\lambda_{\alpha 2}^{\prime} &= 2 m_{H^+}^2 \left(3 - \cos2\beta\right) - 4 m_A^2 , \nn \\
2 \lambda_{\beta 1} v^2 &= 4 m_{H^+}^2 + \left(m_h^2 + m_H^2 - 4 m_{H^+}^2 + 2 m_{H^{++}}^2 + \left(m_H^2 - m_h^2\right)\cos2\alpha \right)\csc^2\beta , \nn \\
- \lambda_{\beta 2}^{\prime} &= 2 m_{H^+}^2 \sin^2\beta - m_A^2 + m_{H^{++}}^2 + \left(m_A^2 - 2 m_{H^+}^2 + m_{H^{++}}^2\right) \csc^2\beta . \nn
\end{align}
The scalar cubic couplings are
\begin{align}
\frac{2}{3} \lambda_{hhh}^{\prime} &= \left(2 m_A^2 - m_h^2 - \left(m_h^2 + 2 m_A^2\right)c_{2\alpha}\right) c_{\alpha} c_{\beta} + 4 m_h^2 c_{\alpha}^3 c_{\beta}^{-1} \\
&+ \sqrt{2} \left(m_A^2 - 3 m_h^2 + \left(m_h^2 + m_A^2\right) c_{2\beta}\right)s_{\alpha}^3 s_{\beta}^{-1} , \nn \\
2 \lambda_{hhH}^{\prime} &= \left(2 m_h^2 + m_H^2\right) c_{\alpha} s_{\alpha} \left(3 - c_{2\beta}\right) \left(c_{\alpha} c_{\beta}^{-1} + \sqrt{2} s_{\beta}^{-1} s_{\alpha}\right) \nn \\
&- m_A^2 c_{\beta} \left(s_{\alpha} \left(3 \sqrt{2} t_{\beta}^{-1} s_{2\alpha} - 1\right) + 3 s_{3\alpha}\right) . \nn
\end{align}

\subsection{Quartet$_1$}
The potential ($V = $ \eqref{eq:VquarZ2} +~\eqref{eq:Vquar1} + $V_{\text{SM}}$) minimization conditions are
\begin{align}
\frac{\mu^2}{v^2} &= \lambda c_{\beta}^2 - \frac{\sqrt{21}}{14} \lambda_1 c_{\beta} s_{\beta} + \frac{1}{42} \left(3 \lambda_{\alpha1} + 2 \lambda_{\alpha2}\right) s_{\beta}^2 , \\
\frac{M^2}{v^2} s_{\beta} &= \frac{\sqrt{21}}{6} \lambda_1 c_{\beta}^3 - \frac{1}{6} \left(3 \lambda_{\alpha 1} + 2 \lambda_{\alpha 2}\right) c_{\beta}^2 s_{\beta} - \frac{1}{63} \left(9 \lambda_{\beta 1} + 5 \lambda_{\beta 2}\right) s_{\beta}^3 . \nn
\end{align}
The rotation to the mass basis is most complicated in this case as there are three singly charged scalars. The charged mass matrix is
\begin{align}
\label{eq:Cmassmatrix}
&\mathcal{M}_{H^+} = \frac{v^2 c_{\beta}^2}{42} \\
&\times \begin{pmatrix}
t_{\beta} \left(7 \sqrt{21} \lambda_1 - \lambda_{\alpha 2} t_{\beta}\right) & 2 \left(\sqrt{7} \lambda_{\alpha 2} t_{\beta} - 7 \sqrt{3} \lambda_1\right) & 21 \lambda_1 + \sqrt{21} \lambda_{\alpha 2} t_{\beta} \\
2 \left(\sqrt{7} \lambda_{\alpha 2} t_{\beta} - 7 \sqrt{3} \lambda_1\right) & 7 \sqrt{21} \lambda_1 t_{\beta}^{-1} + 2 \lambda_{\beta 2} t_{\beta}^2 - 7 \lambda_{\alpha 2} & \tfrac{4 \sqrt{3}}{3} \lambda_{\beta 2} t_{\beta}^2 \\
21 \lambda_1 + \sqrt{21} \lambda_{\alpha 2} t_{\beta} & \tfrac{4 \sqrt{3}}{3} \lambda_{\beta 2} t_{\beta}^2 & 7 \sqrt{21} \lambda_1 t_{\beta}^{-1} + \tfrac{8}{3} \lambda_{\beta 2} t_{\beta}^2 + 7 \lambda_{\alpha 2}
\end{pmatrix} . \nn
\end{align}
The determinant of $\mathcal{M}_{H^+}$ is zero, as required by having a massless Goldstone boson. This also allows us to write the masses of the charged Higgs bosons, $m_{H_{1, 2}^+}$, as
\begin{align}
\label{eq:Cmasses}
2 m_{H_{1, 2}^+} &= \overline{m}_{H^+}^2 \mp \Delta m_{H^+}^2 , \\
\overline{m}_{H^+}^2 &= \text{Tr}\left(\mathcal{M}_{H^+}\right) , \nn \\
\Delta m_{H^+}^4 &= 2 \text{ Tr}\left(\mathcal{M}_{H^+}^2\right) - \text{Tr}\left(\mathcal{M}_{H^+}\right)^2 . \nn
\end{align}
As physical parameters we choose the masses of the $CP$-even Higgs bosons, their mixing angle, the mass of the $CP$-odd Higgs boson, the mass of the doubly charged Higgs boson, and finally $\overline{m}_{H^+}$, which is twice the average of the mass squared of the singly charged Higgs bosons. In terms of these quantities, the Lagrangian parameters are
\begin{align}
28 \lambda^{\prime} &= 6 \left(m_h^2 + m_H^2 - m_A^2\right) + \left(m_h^2 + m_H^2 + 6 m_A^2\right) c_{\beta}^{-2} + \left(m_h^2 - m_H^2\right) c_{2\alpha} \left(6 + c_{\beta}^{-2}\right) ,  \nn \\
\sqrt{7} \lambda^{\prime}_1 &= 2 \sqrt{3} m_A^2 t_{\beta} ,  \\
14 \lambda^{\prime}_{\alpha 1} &= \frac{1}{25 + 24 c_{2\beta}} \left[112 \left(7 m_{H^{++}}^2 + 2 \overline{m}_{H^+}^2 - 2 m_A^2\right) + 42 \left( 3 m_A^2 +  14 m_{H^{++}}^2 + 4 \overline{m}_{H^+}^2\right) c_{2\beta} \right. \nn \\
&\left. - \sqrt{7} \left(m_h^2 - m_H^2\right) \left(136 + 171 c_{2\beta} + 36 c_{4\beta}\right) s_{\beta}^{-1} c_{\beta}^{-1} s_{2\alpha}\right] , \nn \\
\lambda^{\prime}_{\alpha 2} &= \frac{3}{1 + 49 t_{\beta}^{-2}} \left[6 \left(7 m_{H^{++}}^2 + 2 \overline{m}_{H^+}^2\right) - 63 m_A^2 + 7 \left(11 m_A^2 - 7 m_{H^{++}}^2 - 2 \overline{m}_{H^+}^2\right) s_{\beta}^{-2} \right] , \nn \\
4 \lambda^{\prime}_{\beta 1} &= \frac{1}{25 + 24 c_{2\beta}} \left[40 \left(m_{H^{++}}^2 + 35 \overline{m}_{H^+}^2\right) - 1463 m_A^2 - 486 \left(m_h^2 + m_H^2\right) \right. \nn \\
&\left. - 3 \left(63 m_A^2 + 48 \left(m_h^2 + m_H^2\right) - 10 \left(m_{H^{++}}^2 + 14 \overline{m}_{H^+}^2\right)\right) c_{2\beta}\right.  \nn \\
&\left. + 49 \left(26 m_A^2 + 7 \left(m_h^2 + m_H^2\right) - 20 \overline{m}_{H^+}^2\right) s_{\beta}^{-2} \right. \nn \\
&\left. + \left(m_h^2 - m_H^2\right) c_{2\alpha} \left(486 + 144 c_{2\beta} -343 s_{\beta}^{-2}\right) \right] , \nn \\
4 \lambda^{\prime}_{\beta 2} &= \frac{9}{25 + 24 c_{2\beta}} \left[497 m_A^2 - 8 \left(m_{H^{++}}^2 + 35 \overline{m}_{H^+}^2\right) \right. \nn \\
&\left.+ 3 \left(35 m_A^2 - 2 \left(m_{H^{++}}^2 + 14 \overline{m}_{H^+}^2\right)\right) c_{2\beta} + 196 \left( \overline{m}_{H^+}^2 - 2 m_A^2\right) s_{\beta}^{-2}\right] . \nn 
\end{align}
For generic parameters, the splitting between the masses of the singly charged Higgs bosons is
\begin{align}
\label{eq:split}
\frac{\Delta m_{H^+}^4}{\overline{m}_{H^+}^4} &= 1 + \frac{7}{25 + 24 c_{2\beta}} \frac{m_{H^{++}}^2}{\overline{m}_{H^+}^2} \left(2 + \frac{7 m_{H^{++}}^2}{4 \overline{m}_{H^+}^2}\right)  \\
&- \frac{1470 c_{\beta}^2}{\left(4 + 3 c_{2\beta}\right) \left(25 + 24 c_{2\beta}\right)} \frac{m_A^2}{\overline{m}_{H^+}^2}\left(1 + \frac{3}{5} \frac{m_{H^{++}}}{\overline{m}_{H^+}^2}\right) + \frac{147 c_{\beta}^2 \left(155 + 27 c_{2\beta}\right)}{2 \left(4 + 3 c_{2\beta}\right)^2 \left(25 + 24 c_{2\beta}\right)} \frac{m_A^4}{\overline{m}_{H^+}^4} . \nn
\end{align}
We caution that Eq.~\eqref{eq:split} will not work in certain special cases, such as when there are degeneracies in some of the mass parameters. Note however that Eq.~\eqref{eq:Cmasses} will always be correct. Lastly, the cubic couplings are
\begin{align}
\frac{7}{3} \lambda_{hhh}^{\prime} &= m_h^2 \left(4 + 3 c_{2\beta}\right) c_{\beta}^{-1} \left(c_{\alpha}^3 - \sqrt{7} t_{\beta}^{-1} s_{\alpha}^3\right)  \\
&+ m_A^2 \left( c_{\alpha} \left(10 -11 c_{2\alpha}\right) c_{\beta} + c_{\alpha}^3 c_{\beta}^{-1} + \sqrt{7} s_{\alpha} \left(5 c_{2\alpha} - 2 + 7 s_{\beta}^{-2} s_{\alpha}^2\right) s_{\beta}\right) , \nn \\ 
14 \lambda_{hhH}^{\prime} &= \left(2 m_h^2 + m_H^2\right) \left(4 + 3 c_{2\beta}\right) \left(c_{\alpha} c_{\beta}^{-1} + \sqrt{7} s_{\alpha} s_{\beta}^{-1}\right) s_{2\alpha} \nn \\
&+ 3 m_A^2 \left( s_{\alpha} \left(2 c_{\alpha}^2 c_{\beta}^{-1} - 7 \sqrt{7} s_{\beta}^{-1} s_{2\alpha}\right) + c_{\beta} \left(3 s_{\alpha} - 11 s_{3\alpha}\right) + \sqrt{7} \left(3 c_{\alpha} - 5 c_{3\alpha}\right) s_{\beta}\right) . \nn
\end{align}

\subsection{Quartet$_3$}
The potential ($V = $ \eqref{eq:VquarZ2} +~\eqref{eq:Vquar3} + $V_{\text{SM}}$) minimization conditions are
\begin{align}
6 \frac{\mu^2}{v^2} &= 6 \lambda c_{\beta}^2 - 3 \sqrt{3} \lambda_1 c_{\beta} s_{\beta} + \left(\lambda_{\alpha1} + \lambda_{\alpha2}\right) s_{\beta}^2 , \\
6 \sqrt{3} \frac{M^2}{v^2} s_{\beta}^{-2} &= 9 \lambda_1 t_{\beta}^{-3} - 3 \sqrt{3} \left(\lambda_{\alpha1} + \lambda_{\alpha2}\right) t_{\beta}^{-2} - 2 \sqrt{3} \left(\lambda_{\beta1} + \lambda_{\beta2}\right) . \nn
\end{align}
The six quartic couplings can be traded for $m_h$, $m_H$, $m_A$, $m_{H^{+}}$, $m_{H^{++}}$, and $\alpha$, the mixing angle between the $CP$-even Higgs bosons. The mixing angle analogous to $\alpha$ in~\eqref{eq:singmix} for the charged states is $\gamma = - \beta$, and similarly for the $CP$-odd states we have $\tan\delta = - \sqrt{3} \tan\beta$. The quartic couplings are
\begin{align}
4 \lambda^{\prime} &= \left(2 m_A^2 + 3 \left(m_h^2 + m_H^2\right) \right) c_{\beta}^{-2} - 2 \left(m_A^2 + m_h^2 + m_H^2\right) + \left(m_h^2 - m_H^2\right) c_{2\alpha} \left(3 c_{\beta}^{-2} - 2\right), \nn \\
\sqrt{3} \lambda_1^{\prime} &= 2 m_A^2 t_{\beta} , \\
2 \lambda_{\alpha1}^{\prime} &= 6 \left(2 m_{H^+}^2 \left(2 - c_{2\beta}\right) - m_A^2\right) + \sqrt{3} \left(m_H^2 - m_h^2\right) \left(2 - c_{2\beta}\right) s_{\beta}^{-1} c_{\beta}^{-1} s_{2\alpha} , \nn \\
\lambda_{\alpha2}^{\prime} &= 6 \left(m_A^2 - m_{H^+}^2 \left(2 - c_{2\beta}\right)\right) , \nn \\
\frac{4}{3} \lambda_{\beta1}^{\prime} &= 2 \left(m_h^2 + m_H^2 - 2 m_A^2 + 6 m_{H^{++}}^2 - 6 m_{H^+}^2 c_{2\beta}\right)  \nn \\
&+ \left(4 m_A^2 + m_h^2 + m_H^2 - 12 m_{H^+}^2 + 6 m_{H^{++}}^2\right) s_{\beta}^{-2} - 3 \left(m_h^2 - m_H^2\right) c_{2\alpha} \left(2 + s_{\beta}^{-2}\right) , \nn \\
2 \lambda_{\beta2}^{\prime} &= 9 \left(m_A^2 + 2 m_{H^+}^2 c_{2\beta} - 2 m_{H^{++}}^2 + \left(2 m_{H^+}^2 - m_{H^{++}}^2 - m_A^2\right) s_{\beta}^{-2}\right) .\nn
\end{align}
Finally, the mass of the triply-charged Higgs boson is
\begin{equation}
m_{H^{3+}}^2 = \frac{3}{2} m_{H^{++}}^2 + \frac{1}{4} m_A^2 \left(1 - 3 \rho \right) .
\end{equation}
In terms of mass parameters, $c_6$ is
\begin{align}
&c_6 m_h^2 =  \frac{32 m_A^4 s_{\beta} t_{\beta}^2}{3 \left(2 - c_{2\beta}\right)^2 \left[\left(m_h^2 + m_H^2 + \left(m_H^2 - m_h^2\right) c_{2\alpha}\right) s_{\beta} + \sqrt{3} \left(m_H^2 - m_h^2\right) s_{2\alpha} c_{\beta}\right]} .
\end{align}
The cubic couplings are
\begin{align}
&\lambda_{hhh}^{\prime} = 3 m_h^2 \left( c_{\alpha}^3 \left(3 c_{\beta}^{-1} - 2 c_{\beta}\right) - \sqrt{3} s_{\alpha}^3 \left(2 - c_{2\beta}\right) s_{\beta}^{-1}\right) \\
&+ m_A^2 \left(c_{\alpha} \left(4 - 5 c_{2\alpha}\right) c_{\beta} + c_{\alpha}^3 c_{\beta}^{-1} + 3 \sqrt{3} s_{\alpha} \left(c_{2\alpha} + s_{\beta}^{-2} s_{\alpha}^{2}\right) s_{\beta} \right) , \nn \\
&4 \lambda_{hhH}^{\prime} = 2 \left(2 m_h^2 + m_H^2\right) \left(2 - c_{2\beta}\right) \left(c_{\alpha} c_{\beta}^{-1} + \sqrt{3} s_{\alpha} s_{\beta}^{-1}\right) s_{2\alpha} \nn \\
&+ m_A^2 \left(\sqrt{3} c_{\alpha} \left(\left(8 - 12 c_{2\alpha}\right) s_{\beta} - 3 s_{\beta}^{-1}\right) + c_{\beta}^{-1}\left(3 \sqrt{3} c_{3\alpha} t_{\beta}^{-1} + \left(2 + c_{2\beta}\right) s_{\alpha} - \left(4 + 5 c_{2\beta}\right) s_{3\alpha}\right) \right) . \nn
\end{align}
\bibliographystyle{utphys}
\bibliography{SMEFT_H6}

\end{document}